\documentclass[12pt]{article}
\usepackage{authblk}
\usepackage{amsmath}
\usepackage{amssymb}
\usepackage[hmargin=20mm, vmargin=20mm]{geometry}
\usepackage{graphicx}
\usepackage{subfig}
\usepackage{color}
\usepackage[font=scriptsize]{caption}

\begin{document}
\title{Parameter constraints from shadows of
Kerr-Newman-dS black holes with cloud strings and quintessence}
\author{Wenfu Cao$^{1,2}$, 
Wenfang Liu$^{1}$, 
and
Xin Wu$^{1,2}$}
\date{}

\maketitle

\vspace{-10mm}

\begin{center}
{\it
$^1$School of Mathematics, Physics and Statistics, Shanghai
University of Engineering Science, Shanghai 201620, China\\\vspace{1mm}
$^2$Center of Application and Research of Computational Physics,
Shanghai University of Engineering Science, Shanghai 201620,
China\\\vspace{1mm}
}
\end{center}

\vspace{8mm}

\abstract{Shadows of the Kerr-Newman-dS black
hole surrounded by quintessence and a cloud of strings are
investigated. For a spherically
symmetric nonrotating black hole, its shadow is
circular and its size is independent of an observation angle and a
plane on which a circular photon orbit exists.
The shadow sizes are significantly influenced by the parameters involving the cloud of
strings, quintessence parameter, magnitude of quintessential
state parameter, and cosmological constant. The black hole shadows increase with the cloud of strings and negative quintessential
state parameter increasing or the quintessence parameter and cosmological constant
decreasing. When the black hole is spinning
and axially symmetric, the black hole shadow is
dependent on the observation angle and the black hole spin.
The effects of the parameters excluding the spin parameter and  the observation angle
on the sizes of black hole shadows in the rotating case are similar to those in the nonrotating  case. The black hole shadows 
decrease as the black hole spins increase. When the observation angle in the range of 0 and $\pi/2$ is large, the black hole shadow is deformed like the D shape for a high spin, 
but is close to a circle for a low spin. When the observation angle is small, the black hole shadow
seems to be a circle regardless of the high or low spin case.
Based on the Event Horizon Telescope observations of
M87*, the constraint of  the curvature radius is used to constrain
these parameters. For slowly rotating black holes, the allowed
regions of the parameters including the cosmological constant are
given.

}

\vfill{\footnotesize Email: wuxin\_1134@sina.com.}

\maketitle

\newpage
\section{Introduction}

The theory of general relativity  predicts the existence of black
holes in the universe. This prediction has been confirmed through
lots of observation evidences. These evidences include the
detections of the gravitational waves by LIGO [1] and the
observations of the images of supermassive black hole M87* and
SgrA* shadows by the Event Horizon Telescope (EHT) [2].

The shadow of a black hole is a black disk seen by an observer in
the sky when the black hole is illuminated by a light source. This
light source is distributed around the black hole but not between
the observer and the black hole. The computation of black hole
shadow is directly related to the study of photon regions, photon
rings or spheres  outside the event horizon of the black hole. For
a Schwarzschild black hole with mass $M$, the bound photon orbits occur at
$r=3M$, and the critical impact parameter is $\xi_c=3\sqrt{3}M$,
which is the radius of observed photon ring or black hole shadow. In fact,
the circle being the apparent shape of the shadow of a spherically
symmetric black hole was first shown by Synge [3]. Luminet [4]
focused on the appearance of Schwarzschild black hole surrounded
by an accretion disk. Two impact parameters are useful to
determine the apparent positions of the shadow of an axially symmetric
Kerr black hole, which was first investigated by Bardeen [5]. The
shadow of a rotating black hole is no longer circular. The spin of
the black hole leads to the deviation of the shadow from a circle.
There have been many other interesting studies concerning the
shadows of Kerr-Newman black holes [6-8], Kerr-Newman-NUT Black
Holes [9], and black holes surrounded by extra matter sources
[10-14]. The shadows of black holes in modified gravity have also
been considered in numerous publications (see e.g. [15-24]). The
obtained shadow images combined with the observations of  M87* and
Sgr A* shadows are helpful to test theories of gravity and to
understand the geometrical structure of the event horizon and the
parameters of black holes.

The length of a shadow boundary and a local curvature radius are
two characterizations of a black hole shadow [25,26]. The shadow
boundary is a one-dimensional closed or open curve. For a
spherically symmetric nonrotating black hole, the curvature radius
of the black hole shadow is the radius of photon ring. For an
axially symmetric nonvanishing spin black hole, the curvature
radius has maximum and minimum values [27,28]. The minimum and
maximum of the curvature radius determine lower and upper bounds
of the shadow size. Based on the observation of M87*, the black
hole parameters can be constrained via the curvature radius.

Although the theory of general relativity has been confirmed successfully through the observations concerning
the gravitational waves and the images of supermassive black holes,
it cannot completely explain the accelerated
expansion of the universe. This accelerated expansion is caused by  a negative pressure from a gravitationally repulsive energy component.
A possible origin of the negative pressure is quintessence dark energy [29-33]. Dark energy
as an unknown form of energy  accounts for a predominant part of the total energy in the universe.
Quintessence is  a kind of dark energy described by a minimally coupled scalar field.
The state equation of the
dark energy is very associated with the cosmological constant or
vacuum energy. Therefore, the cosmological constant is another possible origin of the accelerated
expansion of the universe. The universe is also thought of as a
collection of extended objects like one-dimensional strings
instead of point particles, termed a
cloud of strings [34,35]. The spacetime structures
of black holes can be affected typically by the cosmological
constant, quintessence and cloud of strings. 
Fathi et al. [36] gave analytical
expressions to the radii of planar and polar spherical photon
orbits around a rotating black hole with quintessential field and
cloud of strings. Critical orbits of particles and photons in the
Schwarzschild black hole with quintessence and string cloud
background spacetimes were investigated by Surya Shankar [37].
Mustafa et al. [38] studied the influence of the cloud of string
parameter and the quintessential parameter  on the radius of the
shadow of the Schwarzschild black hole and the weak defection
angle.  He et al. [39] considered the shadow and photon sphere of
the Schwarzschild  black bole in clouds of strings and
quintessence with static and infalling spherical accretions.
Effect of quintessential dark energy on black hole shadows was
discussed by Singh [40]. Atamurotov et al. [41] investigated the
null geodesics and the shadow cast by the
Kerr-Newman-Kiselev-Letelier black hole for different spacetime
parameters consisting of the quintessence parameter, the cloud of
string parameter, the spin parameter and the charge of the black
hole. The metrics considered in the literature are parts of the
KNdS black hole spacetimes.

Now, we are interested in the study of shadows of
Kerr-Newman-de Sitter (KNdS) black holes with quintessence and a
cloud of strings [29-31]. We particularly focus on the shadows
of KNdS black holes and constraining the black hole parameters
through the curvature radius. For the sake of our purpose,  we introduce the
null geodesic around the KNdS black holes and discuss circular and
spherical photon orbits in Section 2. In Section 3, we obtain the shadow curves
observed by a locally rest observer, and analyze the local
curvature radius for the black hole shadows. Then the parameters
are constrained. Finally, we summarize our main results in Section
4.

\section{Photon motions near KNdS black holes with extra sources}

At first, we introduce a KNdS black hole with quintessence and a
cloud of strings. Then, a Hamiltonian for the description of
photons moving around the black hole is provided. Finally,
circular photon orbits and spherical photon orbits are discussed.

\subsection{KNdS black hole metric}

In Boyer-Lindquist coordinates $(t,r,\theta,\phi)$, the KNdS black
hole surrounded by quintessence and a cloud of strings is
described by the following metric [29]
\begin{eqnarray}
ds^{2} &=&
\frac{dt^2}{\Sigma\Xi^2}(\Delta_{\theta}a^2\sin^{2}\theta-\Delta_{r})+\frac{\Sigma}{\Delta_r}dr^2+
\frac{\Sigma}{\Delta_{\theta}}d\theta^2 \nonumber\\
&&+\frac{2a\sin^{2}\theta}{\Sigma\Xi^2}[\Delta_{r}-\Delta_{\theta}(r^2+a^2)]dtd\phi \nonumber\\
&& + \frac{\sin^{2}\theta}{\Sigma\Xi^2}[\Delta_{\theta}(r^2+a^2)^2
-\Delta_{r}a^2\sin^{2}\theta]d\phi^2,
\end{eqnarray}
where the related notations are defined as
\begin{eqnarray}
 \Sigma &=& r^2+a^2\cos^{2}\theta,
 \\ \nonumber
\Delta_{r}&=&(1-b_{c})r^2+a^2+Q^2-2Mr \\
&&-\frac{\Lambda}{3}r^2(r^2+a^2)-\alpha_{q} r^{1-3\omega_{q}}\\
\Delta_{\theta} &=& 1+\frac{\Lambda}{3}a^2\cos^{2}\theta\\
\Xi &=& 1+\frac{\Lambda}{3}a^2.
\end{eqnarray}
$M$, $Q$ and $a$ stand for the mass, electrical charge and
specific angular momentum of the black hole, respectively. $Q$ and
$a$ are given in the ranges of $|Q|\leq M$ and $|a|\leq M$. In
addition, $\alpha_{q}$ is a positive quintessence parameter, and
$\omega_{q}$ is a quintessential state parameter which satisfies
the condition $-1<\omega_{q}<-1/3$ in a scenario of the
accelerated expansion universe. $b_c$ denotes a positive parameter
measuring the intensity of the cloud of strings [29], and
$\Lambda$ is a positive cosmological constant. In fact, this
metric is  a solution of the Einstein field equation with the
cosmological constant, which can be obtained from the Newman-Janis
transformation of  the nonrotating black hole solution. The total
stress-energy tensor in the nonrotating solution  is a
superposition of three extra sources including the quintessence,
cloud of strings and electromagnetic field. See Refs. [37-39] for
more information on the KNdS spacetime with quintessence and cloud
strings. The speed of light $c$ and the constant of gravity $G$
are taken as geometrical units, $c=G=1$.

Now, let us consider the domains of outer communication. We formally
write $\Delta_r$  as
\begin{equation}
 \Delta_r= (r - r_+) (r - r_-) (r - r_c) f(r),
\end{equation}
where $r_{\pm}$ represent the outer and inner horizons of the black hole, $r_c$ denotes the cosmological horizon [16],
and $f(r)$ is a function of $r$. If the parameters $Q$, $a$, $b_c$, $\alpha_{q}$, $\omega_{q}$ and $\Lambda$ are chosen appropriately,
the equation $\Delta_r=0$ has three real roots $r_{\pm}$ and $r_c$. Figure 1 plots
two parameter spaces for the existence of the real root  $r_c$, where the other parameters are given.
Thus, light rays can reach a rest observer's eyes within the region $r_+\ll r_0\leq r_c$, where $r_0$ denotes the distance of the observer to the black hole.

\subsection{Hamiltonian formulism of photon motions}

The motion of a photon around the black hole can be represented by
the Lagrangian formulism
\begin{equation} \label{eq1}
\mathcal{L}=\frac{1}{2}\frac{ds^2}{d\lambda^2},
\end{equation}
where $\lambda$ is not the proper time but is an affine parameter.
Notice that $\omega_{q}$ and $b_c$ are two dimensionless
parameters. To make the Lagrangian dimensionless, we give scale
transformations to the other quantities: $r\rightarrow rM$,
$t\rightarrow tM$, $a\rightarrow aM$, $Q\rightarrow QM$,
$\Lambda\rightarrow \Lambda/M^2$ and $\alpha_{q}\rightarrow
\alpha_{q}M^{1+3\omega_{q}}$. $\lambda$ is also measured in terms
of the black hole mass, $\lambda\rightarrow \lambda M$. In this
way, the mass factor $M$ is eliminated or becomes 1 in the
Lagrangian.

On the basis of the dimensionless Lagrangian, the photon has a
covariant 4-momentum
\begin{equation}
p_\mu=\frac{\partial \mathcal{L}}{\partial \dot{x}^{\mu}},
\end{equation}
where
$\dot{x}^{\mu}=(\frac{dt}{d\lambda},\frac{dr}{d\lambda},\frac{d\theta}{d\lambda},\frac{d\phi}{d\lambda})$
corresponds to the photon 4-velocity. Because  the coordinates $t$
and $\phi$ do not explicitly appear in the Lagrangian, their
corresponding momenta are conserved. The conserved quantities  are
the photon energy $E$ and angular momentum $L$:
\begin{eqnarray}
p_t &=&\nonumber
\frac{a\sin^{2}\theta}{\Sigma\Xi^2}[\Delta_{r}-\Delta_{\theta}(r^2+a^2)]\dot{\phi}\\
&&\nonumber
+\frac{\dot{t}}{\Sigma\Xi^2}(\Delta_{\theta}a^2\sin^{2}\theta-\Delta_{r})
\\
&=&-E,\\
p_\phi &=&\nonumber
\frac{\sin^{2}\theta}{\Sigma\Xi^2}[\Delta_{\theta}(r^2+a^2)^2-\Delta_{r}a^2\sin^{2}\theta] \dot{\phi}\\
&&\nonumber
+\frac{a\sin^{2}\theta}{\Sigma\Xi^2}
[\Delta_{r}-\Delta_{\theta}(r^2+a^2)]\dot{t}
 \\&=&L.
\end{eqnarray}
Here, $0<E<1$, and $L$ is one of the three possibilities of $L>0$,
$L=0$ and $L<0$. Through a Legendre transformation, the Lagrangian
corresponds to a Hamiltonian formulism
\begin{eqnarray}
\mathcal{H} &=&\nonumber
\frac{E^2}{2}\frac{\Xi^2}{\Sigma}\left[\frac{a^2}{\Delta_{\theta}}\sin^{2}
 \theta-\frac{(r^2+a^2)^2}{\Delta_{r}}\right]\\
&&+\frac{L^2}{2}\frac{\Xi^2}{\Sigma}\left(\frac{1}{\Delta_{\theta}
\sin^{2}\theta}-\frac{a^2}{\Delta_{r}}\right) \nonumber \\
\nonumber
&&-\frac{aEL}{\Sigma}\Xi^2\left(\frac{1}{\Delta_{\theta}}-\frac{r^2+a^2}{\Delta_{r}}\right)\\
&&+\frac{1}{2}\frac{\Delta_r}{\Sigma}p^{2}_{r}+\frac{1}{2}\frac{\Delta_{\theta}}{\Sigma}p^{2}_{\theta}.
\end{eqnarray}
Since the Hamiltonian does not explicitly depend on the affine
parameter $\lambda$, it is a third constant of motion. This
constant is always identical to zero for the null geodesics:
\begin{equation}
\mathcal{H}=0.
\end{equation}

Set a generating function
$S(r,\theta)=S_{r}(r)+S_{\theta}(\theta)$, where $S_{r}$ and
$S_{\theta}$ are functions satisfying the relations $p_r=\partial
S_{r}(r)/\partial r$ and $p_\theta=\partial
S_{\theta}(\theta)/\partial \theta$. Noting Eqs. (11) and
(12), we have the Hamilton-Jacobi equation\footnote{Strictly
speaking, the generating function  for the Hamiltonian (12) with
$E\rightarrow -p_t$ and $L\rightarrow p_\phi$ should be
$S=-\mathcal{H}\lambda-Et+L\phi+S_{r}(r)+S_{\theta}(\theta)$,
where $\mathcal{H}$ is the third constant of motion (12). The
Hamilton-Jacobi equation is obtained by substituting
$p_t=\frac{\partial S}{\partial t}$, $p_r=\frac{\partial
S}{\partial r}$, $p_\theta=\frac{\partial S}{\partial \theta}$ and
$p_\phi=\frac{\partial S}{\partial \phi}$ into the equation
$\frac{\partial S}{\partial \lambda}+\frac{1}{2}g^{\mu\nu}p_\mu
p_\nu=0$, where $g^{\mu\nu}$ denotes the contravariant tensor of
the  metric (1).}
\begin{eqnarray}
0 &=&\nonumber
\frac{E^2}{2}\frac{\Xi^2}{\Sigma}\left[\frac{a^2}{\Delta_{\theta}}
\sin^{2}\theta-\frac{(r^2+a^2)^2}{\Delta_{r}}\right]
\\
&&+\frac{L^2}{2}\frac{\Xi^2}{\Sigma}\left(\frac{1}{\Delta_{\theta}
\sin^{2}\theta}-\frac{a^2}{\Delta_{r}}\right) \nonumber \\
\nonumber
&&-\frac{aEL}{\Sigma}\Xi^2\left(\frac{1}{\Delta_{\theta}}-\frac{r^2+a^2}{\Delta_{r}}\right)
\\
&&+\frac{1}{2}\frac{\Delta_r}{\Sigma}\left(\frac{\partial S_r}
{\partial
r}\right)^2+\frac{1}{2}\frac{\Delta_{\theta}}{\Sigma}\left(\frac{\partial
S_\theta}{\partial \theta}\right)^2.
\end{eqnarray}
This equation has a separation of the variables
\begin{eqnarray}
K &=&\frac{\Xi^2}{\Delta_{\theta}}
(aE\sin\theta-\frac{L}{\sin\theta})^2 +
\Delta_{\theta}\left(\frac{d S_\theta}{d \theta}\right)^2
\nonumber \\ &=& \frac{\Xi^2}{\Delta_{r}}[(E(r^2+a^2)-aL]^2
-\Delta_{r}\left(\frac{d S_r}{d r}\right)^2,
\end{eqnarray}
where $K$ is a Carter constant [42]. Hence, we have two equations
\begin{eqnarray}
&& \frac{dS_{r}(r)}{dr}=\pm\frac{\sqrt{R(r)}}{\Delta_{r}},
\\
&&
\frac{dS_{\theta}(\theta)}{d\theta}=\pm\frac{\sqrt{\Theta(\theta)}}{\Delta_{\theta}},
\end{eqnarray}
where $R(r)$ and $\Theta(\theta)$ are expressed as
\begin{eqnarray}
R(r) &=& \Xi^{2}[(a^2+r^2)E-aL]^{2} -K\Delta_{r},
\\
\Theta(\theta) &=& K\Delta_{\theta}
-\Xi^2(L\csc\theta-aE\sin\theta)^{2}.
\end{eqnarray}
The equations of motion for the null geodesics are
\begin{eqnarray}
\Sigma \dot{r} &=& \pm\sqrt{R(r)},
\\
\Sigma \dot{\theta} &=& \pm\sqrt{\Theta(\theta)}.
\end{eqnarray}
Two equations regarding $\dot{t}$ and $\dot{\phi}$ can be obtained
from Eqs. (9) and (10). They are also parts of the null
geodesic equations, but are not written because they are not used
in this paper.

Several notable points are given here. (\emph{i}) The Carter
constant $K$ is a fourth constant of motion in the Hamiltonian
system (11). Its existence is because the Hamilton-Jacobi
equation (13) allows the separation of variables. Thus, the null
geodesic is integrable and regular. (\emph{ii}) When the photon
gives place to a test particle, the Hamilton-Jacobi equation still
allows the separation of variables and then the Hamiltonian (11)
has a Carter constant nonequal to $K$. (\emph{iii}) When an external
electromagnetic field surrounds the black hole (that is, this electromagnetic field is
included in the Hamiltonian (11)), the Carter constant is not
present for the motion of a charged particle near the black hole,
but it is if the cosmological constant vanishes, as was reported
in Ref. [29].

\subsection{Circular and spherical photon orbits}

Given $R(r)=0$ in Eq. (17), the energy is solved by
\begin{eqnarray}
E_{+}
=\frac{aL}{a^{2}+r^{2}}+\frac{\sqrt{K\Delta_{r}}}{\Xi(a^{2}+r^{2})}.
 \end{eqnarray}
Taking two impact parameters
\begin{eqnarray}
\xi &=& \frac{L}{E},  ~~~~~~ \eta=\frac{K}{E^{2}},
\end{eqnarray}
we define an effective potential
\begin{eqnarray}
V &=&\frac{E_{+}}{E} =
\frac{a\xi}{a^{2}+r^{2}}+\frac{\sqrt{\eta\Delta_{r}}}{\Xi(a^{2}+r^{2})}.
 \end{eqnarray}
Notice that $E_{+}$ in Eq. (23) is a function of $r$ given by
Eq. (21), and $E$ in Eq. (23) is a certain given value of the
energy. In terms of the two impact parameters, Eqs. (17) and
(18) are rewritten as
\begin{eqnarray}
\frac{R(r)}{E^2} &=& \Xi^{2}[(a^2+r^2)-a\xi]^{2}
-\eta\Delta_{r}=\bar{R}(r),
\\
\frac{\Theta(\theta)}{E^2} &=& \eta\Delta_{\theta}
-\Xi^2(\xi\csc\theta-a\sin\theta)^{2}=\bar{\Theta}(\theta).
\end{eqnarray}

Using Eq. (18), Rahaman et al. [43] obtained another effective
potential
\begin{eqnarray}
V_{e}=1-\left(\frac{dr}{d\bar{\lambda}}\right)^2=1-\frac{\bar{R}(r)}{\Sigma^2},
\end{eqnarray}
where $\bar{\lambda}=\lambda/E$.

\subsubsection{Circular photon orbits}

If $\Theta(\theta) =0$ in Eq. (25) for any time $\lambda$, then
$\theta$ remains invariant and photon orbits are always lying on a
certain two-dimensional plane $\theta=\vartheta$. In this case,
$\eta$ is given by
\begin{eqnarray}
\eta_{\vartheta}=\frac{\Xi^2}{\Delta_{\theta}|_{{\theta=\vartheta}}}(\xi\csc\vartheta-a\sin\vartheta)^{2}.
\end{eqnarray}
The motions of photons on the plane are governed by the effective
potential
\begin{eqnarray}
V_{1}=V|_{\eta=\eta_{\vartheta}}, or ~
V_{e1}=V_e|_{\eta=\eta_{\vartheta}}.
\end{eqnarray}

Let us take the parameters $a=0.5$, $Q=0.2$, $b_c=0.01$,
$\alpha_q=0.01$, $\omega_q=-0.35$, $\xi=8.4$, and
$\Lambda=1.02\times 10^{-26}$. Here, such a cosmological constant
is a scaled value, labeled as $\Lambda_{sca}$. It corresponds to a
realistic value of the astrophysical scenario
$\Lambda_{rea}=c^{4}/(M^{2}G^{2})$. If the black hole mass $M$ is
the Sun's mass $M_{\odot}$, we have the realistic value
$\Lambda_{0}=c^{4}/(M_{\odot}^{2}G^{2})={R_{0}}^{-2}=
(M_{\odot}G/c^{2})^{-2}=(1.475km)^{-2}=4.597\times10^{-7}m^{-2}$.
For the supermassive black hole candidate with mass
$M=6.5\times10^{9}M_{\odot}$ in the center of the giant elliptical
galaxy M87, we have $\Lambda_{rea}=\left(M_{\odot}
/M\right)^2\Lambda_{0}=1.088\times10^{-26}m^{-2}$. Thus,
$\Lambda_{sca}=1.02\times10^{-26}M^{-2}$ corresponds to the
realistic cosmological constant
$\Lambda_{rea}=1.11\times10^{-52}m^{-2}$, which was obtained from
the Planck data [44,45]. For simplicity, the subscripts such as
$\emph{sca}$ are dropped in the scaled quantities like
$\Lambda_{sca}$.

 Fig. 2 plots the relation between the radial
distance $r$ and the effective potential $V_1$ or $V_{e1}$ on the
equatorial plane $\vartheta=\pi/2$ in several combinations of the parameters. When $\xi<\xi_{cp}$, the photon
will fall into the black hole; but the photon will scatter to
infinity when $\xi>\xi_{cp}$. When $\xi=\xi_{cp}$, the  photon will wind many times on a circular
orbit. The circular orbit corresponds to the top point (i.e., the
local maximum) of the effective potential $V_1$ or $V_{e1}$.
It is clear that the circular photon orbit satisfies the conditions
\begin{equation}
V_{1}=1, \frac{dV_1}{dr}=0; or ~ V_{e1}=1, \frac{dV_{e1}}{dr}=0.
\end{equation}
The  circular photon orbit is unstable because
\begin{equation}
\frac{d^2V_1}{dr^2}<0, or ~ \frac{d^2V_{e1}}{dr^2}<0.
\end{equation}
The conditions (29) and (30) for the unstable circular photon
orbit are equivalent to the following conditions [29]
\begin{eqnarray}
&&R(r) = 0, ~~~~ \frac{dR(r)}{dr}=0, \\
&& \frac{d^2R(r)}{dr^2}>0.
\end{eqnarray}
The circular orbital radius $r_{cp}$ increases as any one of the parameters
$b_c$, $\alpha_{q}$ and $|\omega_{q}|$ increases in Fig. 2.

In fact, $\xi_{cp}$ and $r_{cp}$ are determined by  Eq. (29) or
(31) when another impact parameter $\eta$ is given according to
Eq. (27). Based on Eq. (31), $\xi_{cp}$ and $\eta_{cp}$ can be
expressed in terms of $r_{cp}$ as
\begin{eqnarray}
\xi_{cp} &=&
a+\frac{r_{cp}}{a}\left(r_{cp}-4\frac{\Delta_r}{\Delta^{'}_r}\right)~~(a\neq0),
 \\
\eta_{cp} &=& 16\Xi^2r_{cp}^2\frac{\Delta_r}{\Delta^{'2}_r},
\end{eqnarray}
where $\Delta^{'}_r=d\Delta_r/dr$. Because of Eq. (27),
$\xi_{cp}$ and $\eta_{cp}$ also satisfy the relation
\begin{eqnarray}
\eta_{cp}=\frac{\Xi^2}{\Delta_{\theta}|_{{\theta=\vartheta}}}(\xi_{cp}
\csc\vartheta-a\sin\vartheta)^{2},
\end{eqnarray}
equivalently,
\begin{eqnarray}
\xi_{cp}=\sin\vartheta
\left(a\sin\vartheta\pm\frac{\sqrt{\eta_{cp}\Delta_{\theta}}}{\Xi}
\right).
\end{eqnarray}
Thus,  $r_{cp}$ can be solved from Eqs. (33), (34) and (36)
for a given value $\vartheta$. Then, $\xi_{cp}$ and $\eta_{cp}$
are determined through Eqs. (33) and (34).

In terms of Eqs. (33)-(35), the radii of circular photon
orbits on the equatorial plane for the Kerr black hole are
expressed as
\begin{eqnarray}
r^{\mp}_{cp}=2\left(1+\cos\left(\frac{2}{3}\arccos(\mp
a)\right)\right),
\end{eqnarray}
where the upper sign ``$-$'' correspond to prograde orbits and the
lower sign ``+'' to retrograde orbits. Then, $\xi_{cp}$ and
$\eta_{cp}$ for the circular photon orbits on the equatorial plane
can be given by Eqs. (33) and (34). When the Kerr black hole
is surrounded by the extra sources, we have no way to provide an
explicit expression of $r_{cp}$. Eqs. (33), (34) and (36)
must be solved through an iterative method, such as the Newtonian
iterative method. In fact, $r_{cp}$ should be solved iteratively,
and has two roots $r^{+}_{cp}$ and $r^{-}_{cp}$ with
$r^{+}_{cp}>r^{-}_{cp}$ for any angle $\vartheta$ in the interval
$0<\vartheta<\pi$.

If $a=0$, the spacetime (1) is spherically symmetric and
$\xi_{cp}$ cannot be given by Eq. (33) but can be given by Eq.
(36). It is expressed as
\begin{eqnarray}
\xi_{cp}=\pm\sin\vartheta \sqrt{\eta_{cp}}.
\end{eqnarray}
In this case, we use Eq. (31) to obtain
\begin{eqnarray}
4\Delta_r=r_{cp}\Delta^{'}_r.
\end{eqnarray}
For the Schwarzschild black hole, Eq. (39) has the solution
$r_{cp}=3$, which corresponds to the radius of circular photon
orbit in the Schwarzschild spacetime. Noting Eq. (34), we have
$\eta_{cp}=27$. Hence, $\xi_{cp}=\pm 3\sqrt{3}$ on the equatorial
plane is obtained from Eq. (38). For the non-Schwarzschild case,
the solution $r_{cp}$ of Eq. (39) is solved by the Newtonian
iterative method. Then, $\eta_{cp}$ is given by Eq. (34), and
$\xi_{cp}$ is obtained from Eq. (38).  It is clear that when the
parameters $b_c$, $Q$, $\Lambda$, $\alpha_q$ and $\omega_q$ are
given, $r_{cp}$ and $\eta_{cp}$ are also determined. However,
$\xi_{cp}$ is varied with a variation of $\vartheta$. Particular
for $\vartheta=\pi/2$, $\xi_{cp}$ has the maximum $\xi^{max}_{cp}=
\sqrt{\eta_{cp}}$.

Clearly, the path of the obtainment of $r_{cp}$, $\xi_{cp}$ and
$\eta_{cp}$ for the rotating case is somewhat unlike that for the
nonrotating case. It is easier to obtain the solutions of
$r_{cp}$, $\xi_{cp}$ and $\eta_{cp}$ for the nonrotating case than
those for  the rotating case.

\subsubsection{Spherical photon orbits}

As is demonstrated above, the effective potential (28) governs
the radial motion of photons on the two-dimensional plane. If
$\Theta$ is not always identical to zero for any time $\lambda$,
then $\theta$ varies in the range $0<\theta<\pi$ with time
$\lambda$ in Eq. (25). Note that $\eta$ satisfying Eq. (25),
i.e. $\theta=\vartheta$ as a solution of $\Theta(\theta) =0$, is
possible at some time, but is impossible at any other times.
Without loss of generality, $\theta$ satisfying the condition
$\Theta(\theta)\geq 0$ is arbitrarily given in the range
$0<\theta<\pi$. In this case, the motion of photons is not lying
on the two-dimensional plane but is lying  in the
three-dimensional space. Eq. (23) or (26) is still the
effective potential in the three-dimensional space, labeled as
$V_2$ or $V_{e2}$. The top point of the effective potential $V_2$
or $V_{e2}$ in Fig. 2(b) corresponds to a spherical photon orbit.
The conditions for the existence of spherical photon orbit are
still the same as Eqs. (29) and (30) (or Eqs. (31) and
(32)) for the existence of circular photon orbit. An explicit
difference between the spherical photon orbit and the circular
photon orbit is that Eq. (27) is not satisfied for the spherical
photon orbit, whereas it is for the circular photon orbit. The two
impact parameters $\xi_{sp}$ and $\eta_{sp}$ for the spherical
photon orbit are consistent with the expressions of $\xi_{cp}$ and
$\eta_{cp}$ for the circular photon orbit in Eqs. (33) and
(34), but do not satisfy Eq. (36). The values of $\xi_{cp}$
are based on $a\neq0$. If $a=0$, then the spacetime (1) is
spherically symmetric, and the above-mentioned circular photon
orbits are present. $\xi_{cp}$ is given in Eq. (38) rather than
Eq. (33).

Using Eq. (32), we can know that for the case of $a\neq0$,
$r_{sp}$ is constrained in the range $r^{-}_{cp}\leq r_{sp}\leq
r^{+}_{cp}$, where $r^{-}_{cp}$ and $r^{+}_{cp}$ are the radii of
circular photon orbits on the plane $\theta=\vartheta$. Note that
$\vartheta$ may not be $\pi/2$. As the spherical radius $r_{sp}$
ranges from $r^{-}_{cp}$ to $r^{+}_{cp}$, an infinite number of
points $(\xi_{sp}, \eta_{sp})$ are obtained from Eqs. (33) and
(34).

\section{Parameter constraints based on black hole shadows}

A black hole shadow is observed by an observer in a zero angular
moment observer (ZAMO) reference frame [5]. Then a local
curvature radius for the boundary of black hole shadow is
discussed. Finally, the constraint of  curvature radius is used to
constrain the parameters.

\subsection{Black hole shadows}

The above-mentioned points $(\xi_{sp}, \eta_{sp})$ from the
spherical photon orbits are used to study the black hole shadows.

In order to obtain an image of the black hole, we introduce
celestial coordinates in an observer's sky. Assume that the static
observer locally stays at point $(r_{0},\theta_{0})$ in the ZAMO
reference frame, where the observer can determine the image
points. The observer basis $\{\hat{e}_{t}, \hat{e}_{r}, \hat{e}_{\theta},
\hat{e}_{\phi}\}$ can be expressed in terms of the coordinate basis
$\{\partial_{t}, \partial_{r},
\partial_{\theta},
\partial_{\phi}\}$ as [46-48]
\begin{eqnarray}
&& \hat{e}_{(\mu)}=\hat{e}^{\nu}_{(\mu)}\partial_{\nu},
\end{eqnarray}
where $\hat{e}^{\nu}_{(\mu)}$ is a transform matrix satisfying the
relation
$g_{\mu\nu}e^{\mu}_{\alpha}e^{\nu}_{\beta}=\eta_{\alpha\beta}$
with $\eta_{\alpha\beta}$ being the Minkowski metric. In general,
it is convenient to choose the observer located in the frame
\begin{eqnarray}\label{et}
  \hat{e}_{(t)} & = & \sqrt{\frac{g_{\phi \phi}}{g_{t \phi}^2 - g_{t t} g_{\phi
  \phi}}} \left( \partial_t - \frac{g_{t \phi}}{g_{\phi \phi}} \partial_{\phi}
  \right),\label{et}\\
  \hat{e}_{(r)} & = & \frac{1}{\sqrt{g_{r r}}} \partial_r,\\
  \hat{e}_{(\theta)} & = & \frac{1}{\sqrt{g_{\theta \theta}}} \partial_{\theta},\\
  \hat{e}_{(\phi)} & = & \frac{1}{\sqrt{g_{\phi \phi}}} \partial_{\phi},
\end{eqnarray}
Since $\hat{e}_{(t)}\cdot \partial_{\phi}=0$, the observer in this local rest
frame has zero angular momentum at infinity. In this sense, the
frame is called the ZAMO reference frame, representing the zero
angular momentum observer.

The four-momentum $p^{\mu}$ of a photon by its projection  onto
$\hat{e}_{\mu}$ is locally measured by
\begin{eqnarray}
&& p^{(t)}=-p_{\mu}\hat{e}^{\mu}_{(t)},
\\
&& p^{(i)}=p_{\mu}\hat{e}^{\mu}_{(i)} ~~~~(i=r, \theta, \phi).
\end{eqnarray}
On the basis of Eqs. (41-44), the four-momentum $p^{\mu}$ can be
rewritten as
\begin{eqnarray}
&&
p^{(t)}=\zeta E-\gamma L,
\\
&&
p^{(r)}=\frac{1}{\sqrt{g_{rr}}}p_{r},
\\
&&
p^{(\theta)}=\frac{1}{\sqrt{g_{\theta\theta}}}p_{\theta},
\\
&& p^{(\phi)}=\frac{1}{\sqrt{g_{\phi\phi}}}p_{\phi},
\end{eqnarray}
where $\zeta= \hat{e}_{(t)}^t$ and $\gamma=\hat{e}_{(t)}^\phi$.

Suppose that the observer has a radial distance $r_{0}$ in Fig. 3.
The inclination angle between the symmetrical axis of the black
hole and the direction to the observer is $\theta_{0}$. The
3-vector $\vec{p}$ is the photon's linear momentum with three
components $p^{(r)}, p^{(\theta)}$ and $p^{(\phi)}$ in the orthonormal
basis $\{\hat{e}_{t}, \hat{e}_{r}, \hat{e}_{\theta}, \hat{e}_{\phi}\}$ [46-48]:
\begin{eqnarray}
&&
\vec{p}=p^{(r)}\hat{e}_{r}+p^{(\theta)}\hat{e}_{\theta}+p^{(\phi)}\hat{e}_{\phi}.
\end{eqnarray}
Note that $(p^{(t)})^2=\mathbf{p}^2=(p^{(r)})^2+(p^{(\theta)})^2+(p^{(\phi)})^2$.
The observation angles $(\alpha,\beta)$ are introduced by
\begin{eqnarray}
&&
p^{(r)}=|\vec{p}|\cos\alpha\cos\beta,
\\
&&
p^{(\theta)}=|\vec{p}|\sin\alpha,
\\
&&
p^{(\phi)}=|\vec{p}|\cos\alpha\sin\beta.
\end{eqnarray}
Thus, we have
\begin{eqnarray}
  \sin \alpha& =&  \frac{p^{(\theta)}}{p^{(t)}}=\pm\frac{1}{(\zeta-\gamma \xi)\triangle_{\theta}}
\sqrt{\frac{\Theta}{g_{\theta\theta}}} ,\label{angles1} \\
    \tan \beta &=& \frac{p^{(\phi)}}{p^{(r)}}=\frac{\xi \sqrt{g_{rr}}\Delta_r}
    {\sqrt{g_{\phi\phi}R}},\label{angles2}.
\end{eqnarray}

An image point is  described in celestial coordinates $(x, y)$ [48] by
\begin{equation}
  x \equiv - r_0 \beta, \quad y \equiv r_0\alpha. \label{xy}
\end{equation}
Because
\begin{eqnarray}\label{sima}
\sin \alpha &\rightarrow& \frac{\sqrt{1-b_{c}-\frac{\Lambda}{3}r_{0}^{2}-\alpha_{q}r_{0}^{-1-3\omega_{q}}}\sqrt{\eta_{p}-(\xi_{p}\csc\theta_{0}-a\sin\theta_{0})^{2}}}{r_{0}},
\\
\tan\beta &\rightarrow& \frac{\xi_{p}\sqrt{1-b_{c}-\frac{\Lambda}{3}r_{0}^{2}-\alpha_{q}r_{0}^{-1-3\omega_{q}}}}{r_{0}\sin\theta_{0}},
\end{eqnarray}
$x$ and $y$ satisfy the relation
\begin{eqnarray}\label{sima}
x^{2}+y^{2}&=& (1-b_{c}-\frac{\Lambda}{3}r_{0}^{2}-\alpha_{q}r_{0}^{-1-3\omega_{q}})(\eta_{p}+2a\xi_{p}-a^{2}\sin^{2}\theta_{0}).
\end{eqnarray}
$r_0=10^{10}$ is so large that $\sin \alpha$ and $\tan\beta$ are extremely small in Eqs. (58) and (59).
In this case, $\sin \alpha\approx \alpha$ and $\tan\beta\approx\beta$.

For the nonrotating case of $a=0$, Eq. (60) becomes
\begin{equation}
x^{2}+y^{2}= (1-b_{c}-\frac{\Lambda}{3}r_{0}^{2}-\alpha_{q}r_{0}^{-1-3\omega_{q}})\eta_{p}=R^{2}_{sh}.
\end{equation}
Obviously, the black hole shadow is a standard circle with the radius
$R_{sh}$.
The Schwarzschild black hole shadow has its radius $R_{sh}=3\sqrt{3}$. The shadow
size is independent of the angles $\vartheta$ and $\theta_{0}$.
The result is still present when the extra sources such as the
quintessence, cloud strings, cosmological constant and black hole
charge are included in the Schwarzschild spacetime. This is
because $r_{cp}$ is given by Eq. (39) that does not depend on
the angles $\vartheta$ and $\theta_{0}$, and $\eta_{cp}$ is
obtained from Eq. (34) that does not contain the angles
$\vartheta$ and $\theta_{0}$. In fact, $r_{cp}$ depends on the
parameters $b_{c}$, $\alpha_{q}$, $\omega_{q}$ and $\Lambda$.
$\eta_{p}$ or $R_{sh}$ is also determined by these parameters.
Table 1 lists the values of $r_{cp}$ and $R_{sh}$ for several different combinations of the
parameters. The radius of circular photon orbit increases as each
of the parameters $b_{c}$, $\alpha_{q}$ and $|\omega_{q}|$ increases.
However, a variation of $\Lambda$ does not affect the radius of circular photon orbit
because $r_{cp}$ in Eq. (39) is independent of $\Lambda$.
The shadow $R_{sh}$ increases as either the parameter $b_{c}$ or $\omega_{q}$ increases. This result is due to the
increase of the black hole gravitational field. On the contrary, as the parameters $\alpha_{q}$ and $\Lambda$ increase,
the gravitational field of the black hole is weakened, and the size of the black hole
shadow decreases.

For the rotating case $a\neq0$, the celestial coordinates $(x, y)$
are obtained unlike those for the nonrotating case. If the
observer stays at the equatorial plane $\theta_{0}=\pi/2$, the
celestial coordinates have the relation
\begin{eqnarray}
\left(x+a\sqrt{1-b_{c}-\frac{\Lambda}{3}r_{0}^{2}-\alpha_{q}r_{0}^{-1-3\omega_{q}}}\right)^{2}+y^{2}
=\left(1-b_{c}-\frac{\Lambda}{3}r_{0}^{2}-\alpha_{q}r_{0}^{-1-3\omega_{q}}\right)\eta_{p}.
\end{eqnarray}
Notice that $\eta_p$ in Eq. (62) is unlike that in Eq. (61).
$\eta_p$ in Eq. (61) is dependent on the radius of spherical
photon orbits (i.e. the radius of circular photon orbits) and has
only one invariant value. Therefore, the shadow for the
nonrotating black hole is a standard circle. However, there are
two photon circular orbits on a plane in Eq. (62), and the radii
of spherical photon orbits are arbitrarily given between the two
radii of circular photon orbits. That is, $\eta_p$ in Eq. (62)
is varied. In this situation, Eq. (62) cannot be thought of as a
circle for the rotating black hole.

For the Kerr spacetime, the radii of photon circular orbits on the
equatorial plane are $r^{\mp}_{cp}$ in Eq. (37). As the
photon spherical radii $r_{p}$ range from $r^{-}_{cp}$ to
$r^{+}_{cp}$, $\xi^{\mp}_{p}$ and $\eta^{\mp}_{p}$ can be given by
Eqs. (33) and (34). In this way, all celestial coordinate
points on the shadow are obtained. If the observer is not located
on the equatorial plane (i.e.  $\theta_{0}\neq\pi/2$), we can
still use this method but have to discard imaginary points during
the calculations. When $\theta_{0}$ is consistent with
$\vartheta$, the celestial coordinates are no longer determined by the
spherical photon orbits, whose radii are given in the range
$(r^{-}_{cp}, r^{+}_{cp})$ in Eq. (37). They should be
determined by  the spherical photon orbits, whose radii  are
arbitrarily given in the range $(r^{-}_{cp}, r^{+}_{cp})$. Here,
$r^{-}_{cp}$ and $r^{+}_{cp}$ are the radii of  circular photon
orbits for satisfying Eqs. (33), (34) and (36) on the planes
$\vartheta$. In this way, the smallest radius of circular photon orbit $r_D$ and the largest radius of circular photon
orbit $r_R$ are provided. The circular photon orbits with the radii $r_D$ and $r_R$ are termed critical circular photon orbits.
The method for calculating $r_D$ and $r_R$ is termed Method 1. Another path for the obtainment of $r_D$
and $r_R$ is to solve the equation $y(r_{cp})=0$, where $r_{cp}$
corresponds to the radii of unstable circular photon orbits on the
planes $\theta_{0}$. In fact, the celestial
coordinate $y$ is identical to zero for the critical circular photon orbits when $\eta_{cp}$ in Eq. (35) is substituted into
Eq. (55). This fact is why the radii of circular photon orbits
are obtained by solving  the equation $y(r_{cp})=0$. The method for calculating $r_D$ and $r_R$ is termed Method 2.
Fig. 4 (a)-(d) shows that the two methods
give the relations between the critical circular orbital radii and $\theta_{0}$ for two spin cases in the Kerr spacetime.
Method 1 (Fig. 4 (a) and (c) in the left-hand side) and Method 2 (Fig. 4 (b) and (d) in the right-hand side)
give the same numerical results to the critical circular orbital
radii $r_D$ and $r_R$ on the planes $\vartheta=\theta_{0}$. This result is also suitable for various observational angles $\theta_{0}$,
as shown in Table 2. When the spin $a$ is given,  $r_D$ and $r_R$ decrease with the observational angle $\theta_{0}$ increasing from 0 to $\pi/2$.
When the observational angle $\theta_{0}$ is given,  $r_D$ and $r_R$ for the slower spin $a=0.5$ are larger than those for  the higher spin $a=0.9$.
It is also shown via  Table 2 that $r_D$ and $r_R$ are symmetric with respect to $\theta_{0}=\pi/2$. Due to $r_D=r_R$ for $\theta_{0}=0,\pi$,
the shadows are circular in this case. When the additional sources $Q=0.2$, $b_{c}=0.01$, $\alpha_{q}=0.01$, $\omega_{q}=-0.35$ and $\Lambda=1.02\times 10^{-26}$
are included in Fig. 4(c), Method 1 (Fig. 4(e) in the left-hand side) and Method 2 (Fig. 4(f) in the right-hand side)
also give the same numerical results to $r_D$ and $r_R$ in the KNdS spacetime. The additional sources slightly affect $r_D$ and $r_R$  in the KNdS spacetime,
as compared with those in the Kerr case. For a given observational angle $\theta_{0}$ in Table 2, $r_D$ and $r_R$ in the
former case are slightly larger  than or approximately equal to those in
the latter one. The points $(x_{D},0)$ and $(x_{R},0)$ of the Kerr black hole shadow are plotted in Fig. 5,
and they are determined by the prograde and retrograde photon orbits
in the plane $\vartheta=\theta_{0}$. In practice, the black hole
shadow is made of the two points and other points associated with
unstable spherical photon orbits between the radii of two circular
photon orbits in the plane $\vartheta=\theta_{0}$.

The above demonstrations introduce the two methods for
computing the Kerr black hole shadows. These  methods are
still suitable for the  computation of black hole shadows when the
quintessence, cloud strings, cosmological constant and black hole
charge are included in the Kerr spacetime.

\subsection{Local curvature radius}

The boundary of a black hole shadow can describe some  properties
of the black hole. It is a one-dimensional closed curve in the
celestial coordinates.  From the viewpoint of differential
geometry, the curve has its length and local curvature radius.
The curvature radius is written in Ref. [57] as
\begin{eqnarray}
&& \Re=\left|\frac{(x'(r_{p})^{2}+y'(r_{p})^{2})^{(3/2)}}
{x'(r_{p})y''(r_{p})-x''(r_{p})y'(r_{p})}\right|.
\end{eqnarray}

Utilizing the symmetry of black hole shadow, the authors of [54]
discussed several characteristic points along the boundary curve
of the Kerr black hole shadow in Fig. 5. These points are $D$,
$R$, $B$ and $T$. The Kerr black hole parameters can be
constrained in terms of the characteristic points and curvature
radius. The points $T$ and $B$ are determined by
\begin{eqnarray}
\frac{dy}{dr_p}=0.
\end{eqnarray}
The points $D$ and $R$ are governed by
\begin{eqnarray}
 y=0.
\end{eqnarray}
Since the shadow curve has the $Z2$ symmetry, the vertical
diameter $\Delta y$ of the shadow is
\begin{eqnarray}
&& \Delta y=2y_{T}.
\end{eqnarray}
The horizontal diameter is given by
\begin{eqnarray}
&& \Delta x=|x_{D}-x_{R}|.
\end{eqnarray}

Similar to the Kerr black hole shadow, the KNdS black hole shadows can also be considered.
Figs. 6-9 plot the KNdS black hole shadows for different combinations of the parameters.
When the black hole spin $a=0.98$ is high in Fig. 6, the black hole shadows are not circles but look like ``D".
The shadows increase with the parameters $b_c$ and $\omega_{q}$ increasing, but decrease with the parameters $\alpha_{q}$ and $\Lambda$ increasing.
When the black hole spin $a=0.5$ is moderate in Fig. 7, the black hole shadows are approximately circles. The effects of the parameters
on the shadows in this case are similar to those in the high spin case. When the black hole spin $a=0.1$ is low in Fig. 8,
the black hole shadows are more circular. The effects of the parameters
on the shadows in the low spin case are similar to those in the high or moderate spin case. Fig. 9 describes the black hole shadows in four spacetimes
involving the Kerr, Kerr-de Sitter (KdS), KNdS (with $b_c=\alpha_q=0$) and KNdS (with $b_c\neq0$ and $\alpha_q\neq0$) spacetimes for three spins $a=0.5, 0.8, 0.94$. The three spins
are considered because the related  constraints are suitable for  the black hole spin values of $a$ from 0.5 to 0.94 [26]. These results
show that  the black hole shadows seem to be circles whether the spin is small or large.
When the spin is given, the three black hole shadows in Fig. 9 (a)-(c) and Table 6 are indistinguishable.
The result is consistent with that of [28] about the relation between the KdS black hole
shadow and the Kerr black hole shadow in the case of $\Lambda=1.02\times10^{-26}$. In addition,
the black hole shadows decrease when the spins increase. Here the observation angle is $\theta_0=17^{\circ}$, which is the angle between the approaching jet from the
central radio source in M87* and the line of sight [49].
Because a highly charged dilaton black hole is ruled out
by the measurements of the EHT [6], smaller
values are given to the black hole charges in our work. The black hole shadow is more circular for the large spin  in Fig. 9 than that for the large spin in Fig. 6.
It is because the observation angle $\theta_0=17^{\circ}$ in Fig. 9 is smaller than the observation angle $\theta_0=90^{\circ}$ in Fig. 6.
Although the observation angle $\theta_0=90^{\circ}$  is still larger in Fig. 8, the spin $a=0.1$ is so smaller that the black hole shadow is close to a circle. 

 More details on the characteristics of the black hole
shadows in Figs. 6-9 are listed in Tables 3-6. Here the curvature radii
at the points $T$, $D$ and $R$, which  respectively correspond to $\Re_{T}$,
$\Re_{D}$ and $\Re_{R}$, are computed by the gravitational radius. The gravitational radius from the EHT observations [2,25] is expressed as
\begin{eqnarray}
\theta_{g}=\frac{GM}{c^{2}l}\approx3.8\mu as,
\end{eqnarray}
where $l$ is the distance from the observer to the black hole, i.e. $l=r_{0}$. For the larger observation angle in Tables 3-5,  $\Re_{T}$,
$\Re_{D}$ and $\Re_{R}$ have typical differences for the high spins, but small differences for the low spins.
The shadow size increases as either the parameter $b_{c}$ or $\omega_{q}$ increases. This result is due to the increase of the black hole gravitational field.
On the contrary,  the gravitational field of the
black hole is weakened and the size of the black hole shadow decreases as the parameters $\alpha_{q}$ and $\Lambda$ increase. For the smaller observation angle in Table 6,
the curvature radii at the characteristic points $\Re_{T}$, $\Re_{D}$ and $\Re_{R}$
have small changes with the increases or decreases of these parameters even if the spin is large.
Consequently, the shadows remain nearly circular. 

\subsection{Constraints of the parameters}\label{sec4}

Based on the EHT observations of M87*, the radius of the
shadow is constrained in the range
\begin{eqnarray}
4.31M \leq r_{sh,A} \leq 6.08M.
\end{eqnarray}
Equivalently, the constraint to  the curvature radius was given in
[27-28] by
\begin{eqnarray}
4.31M \leq \Re \leq 6.08M.
\end{eqnarray}

The curvature radius may have a maximum value and a minimum value.
It has two local maximum values at $r_{p}=r_D$ and $r_{p}=r_R$.
The local maximum at $r_{p}=r_D$ is larger than that at
$r_{p}=r_R$. Hence we have $\Re_{max}=\Re(r_D)$. The minimum
curvature radius $\Re_{min}=\Re(r_T)$ corresponds to the well of
these curves. Moreover, $\Re_{min}$ and $\Re_{max}$ give lower and upper bounds
to the size of the shadow. In other words, $\Re_{min}$ should not
decrease below 4.31M and $\Re_{max}$ should not increase beyond
6.08M.

Finally, we utilize the constraint of the curvature radius to restrict
the parameters. The allowed region of the cloud strings
$b_c$ in Fig. 10(a) is down the red curve corresponding to the
maximum curvature radius $\Re_{max}$ when the black hole spin $a$
ranges from 0 to 1 and the other parameters are given.
Similarly, the allowed region of the  quintessence parameter
$\alpha_{q}$ in Fig. 10(b) and the cosmological constant
$\Lambda$ in Fig. 10(d) is down the blue curve corresponding to the
minimum curvature radius $\Re_{min}$ when the black hole spin $a$
ranges from 0 to 1 and the other parameters are given.
However, the allowed region of the quintessential state parameter
$\omega_{q}$ in Fig. 10(c) is the upper region of the blue curve.
For any spin $a\in[0,1]$, the allowed regions of the parameters
are $0\leq b_c<0.15$ in Fig. 10(a), $0\leq\alpha_{q}<0.18$ in Fig. 10(b),
$-0.376<\omega_{q}<-1/3$ in Fig. 10(c), and $0\leq\Lambda<5\times10^{-21}$ in Fig. 10(d).

\section{Conclusions}

In this paper, we focus on the motion of photons around the KNdS
black hole surrounded by quintessence and a cloud of strings.

Due to the existence of the Carter constant, unstable circular photon
orbits on a two-dimensional plane not limited to the equatorial
plane and unstable spherical photon orbits in the
three-dimensional space can be present. The conditions for the
existence of circular photon orbits are basically consistent with
those for the existence of spherical photon orbits. However, only
one difference between them is only that the angle $\theta$ always
remains invariant for the circular photon orbits and is varied
with time for the spherical photon orbits. The two impact
parameters can be determined by these circular photon orbits and
spherical photon orbits. The radius of circular photon orbit
increases as each of the parameters involving the cloud of strings
$b_{c}$, quintessence parameter $\alpha_{q}$, and quintessential
state parameter $|\omega_{q}|$ increases. However, a small variation of the
cosmological constant $\Lambda$ does not typically affect the circular photon orbit radius.

For the nonrotating case with the spherical symmetry, the black
hole shadows are circular and their sizes are independent of the
observation angles and the planes on which circular  photon orbits
exist. For the rotating case with the axialsymmetry, the black hole shadow is
dependent on the observation angle and the black hole spin parameter. In both cases, small changes of the parameters
excluding the spin parameter and the observation angle
exert similar influences on the sizes of black hole shadows. That is, the black hole shadows increase
with the cloud of strings and negative quintessential
state parameter increasing, whereas decrease with the quintessence parameter and cosmological constant
increasing. The black hole shadows also decrease as the black hole spins get larger. When the observation angle in the range of 0 and $\pi/2$ is large,
the black hole shadow
looks like the D shape for a high spin, but is close to a circle for a low spin. When the observation angle is small, the black hole shadow
seems to be a circle regardless of the high or low spin case. 

Based on the EHT observations of M87*, the constraint of  the
curvature radius is used to constrain the parameters. For any spin
$a\in[0,1]$, the allowed regions of the parameters are $0\leq
b_c<0.15$, $0\leq\alpha_{q}<0.18$, $-0.376<\omega_{q}<-1/3$, and
$0\leq\Lambda<5\times10^{-21}$.

\section*{Acknowledgments}

The authors are very grateful to referees for useful suggestions.
This research has been supported by the National
Natural Science Foundation of China (Grant No. 11973020), and the
Natural Science Foundation of Guangxi (Grant No.
2019GXNSFDA245019).

\textbf{Conflict of Interest}

The authors declare that there is no conflict of interest.

\begin{table*}[htbp]
\centering \caption{Values of $r_{cp}$ ($M$) and $R_{sh}$ ($\mu as$)
of circular photon orbits on the equatorial plane for each of the parameters $b_{c}$, $\alpha_{q}$,
$\omega_{q}$ and $\Lambda$ in the nonrotating case.  The black hole charge is $Q=0.2$,
and the other parameters are given as follows:
(a) $\alpha_q=0.01$, $\omega_q=-0.35$ and $\Lambda=1.02\times10^{-26}$;
(b) $b_{c}=0.01$, $\omega_q=-0.35$ and $\Lambda=1.02\times10^{-26}$;
(c) $b_{c}=0.01$, $\alpha_q=0.01$ and $\Lambda=1.02\times10^{-26}$;
(d) $b_{c}=0.01$, $\alpha_q=0.01$ and $\omega_q=-0.35$. 
\label{tab1}}
\begin{tabular}{ccccccccccc}
\hline
  &$b_{c}$&0&0.05&0.1\\
  (a) &$r_{cp}$&3.004&3.165&3.345\\
   &$R_{sh}$&$19.61$&$20.65$&$21.80$\\
\hline
   &$\alpha_{q}$&$0$&$0.1$&$0.15$\\
  (b) &$r_{cp}$&3.003&3.35&3.57\\
   &$R_{sh}$&$19.81$&$19.39$&$18.64$\\
\hline
   &$\omega_{q}$&-0.35&-0.36&-0.37\\
  (c) &$r_{cp}$&3.035&3.036&3.037\\
   &$R_{sh}$&$19.81$&$19.49$&$18.87$\\
\hline
   &$\Lambda$&$10^{-22}$&$10^{-21}$&$10^{-20}$\\
  (d) &$r_{cp}$&3.036&3.036&3.036\\
   &$R_{sh}$&$19.78$&$19.46$&$16.00$\\
\hline
\end{tabular}
\end{table*}

\begin{table*}[htbp]
\centering \caption{The critical circular photon orbital radii
$r_{D}$ ($M$) and $r_{R}$ ($M$) in Fig. 4.}\label{tab2}
\begin{tabular}{ccccccccc}
\hline
    &$\theta_{0}$&$0^{\circ}$&$45^{\circ}$&$90^{\circ}$&$135^{\circ}$&$180^{\circ}$\\
   Fig. 4(a)  & $r_{D}$  &2.88&2.48&2.34&2.48&2.88\\
      & $r_{R}$     &2.88&3.34&3.53&3.34&2.88\\
\hline
  &$\vartheta$&$0^{\circ}$&$45^{\circ}$&$90^{\circ}$&$135^{\circ}$&$180^{\circ}$\\
   Fig. 4(b) & $r_{D}$  &2.88&2.48&2.34&2.48&2.88\\
   & $r_{R}$     &2.88&3.34&3.53&3.34&2.88\\
\hline
  &$\theta_{0}$&$0^{\circ}$&$45^{\circ}$&$90^{\circ}$&$135^{\circ}$&$180^{\circ}$\\
  Fig. 4(c) & $r_{D}$  &2.55&1.73&1.5&1.73&2.55\\
     & $r_{R}$     &2.55&3.52&3.91&3.52&2.55\\
\hline
   &$\vartheta$&$0^{\circ}$&$45^{\circ}$&$90^{\circ}$&$135^{\circ}$&$180^{\circ}$\\
    Fig. 4(d) & $r_{D}$  &2.55&1.73&1.5&1.73&2.55\\
     & $r_{R}$     &2.55&3.52&3.91&3.52&2.55\\
\hline
  &$\vartheta$&$0^{\circ}$&$45^{\circ}$&$90^{\circ}$&$135^{\circ}$&$180^{\circ}$\\
   Fig. 4(e) & $r_{D}$  &2.57&1.73&1.56&1.73&2.57\\
   &   & $r_{R}$     &2.57&3.56&3.96&3.56&2.57\\
\hline
  &$\vartheta$&$0^{\circ}$&$45^{\circ}$&$90^{\circ}$&$135^{\circ}$&$180^{\circ}$\\
   Fig. 4(f) & $r_{D}$  &2.57&1.73&1.56&1.73&2.57\\
  & $r_{R}$     &2.57&3.56&3.96&3.56&2.57\\
\hline
\end{tabular}
\end{table*}

\begin{table*}[htbp]
\centering \caption{Characteristics of the black hole
shadows in Fig. 6.
The horizontal diameter (HD) $(\mu as)$ is $\Delta x$, and the
vertical diameter (VD) $(\mu as)$ is $\Delta y$. The gravitational radius is
$\theta_{g}\approx3.8\mu as$.  \label{tab3}}
\begin{tabular}{ccccccccc}
\hline
   & Fig. 6(a)& $b_{c}$ &0&0.05&0.1&\\
   & HD & $\Delta x$ &34.61&37.38&40.02&\\
   & VD & $\Delta y$ &39.22&41.30&43.61&\\
   && $\Re_{D}$      &75.68&38.99&33.81&\\
   && $\Re_{T}$      &17.45&18.51&19.63&\\
   && $\Re_{R}$      &20.12&21.38&22.33&\\
\hline
   & Fig. 6(b)& $\alpha_{q}$ &0&0.1&0.15&\\
   & HD & $\Delta x$  &34.96&35.61&34.6&\\
   & VD & $\Delta y$  &39.62&38.79&37.28&\\
   & & $\Re_{D}$       &76.49&29.99&25.87&\\
   &   & $\Re_{T}$     &17.69&17.49&17.01&\\
   &   & $\Re_{R}$     &20.33&19.86&19.07&\\
\hline
   & Fig. 6 (c)& $\omega_{q}$ &-0.35&-0.36&-0.37&\\
   & HD & $\Delta x$ &35.22&34.66&33.49&\\
   & VD & $\Delta y$  &39.62&38.99&37.67&\\
   & & $\Re_{D}$  &56.78&55.85&53.95&\\
   &   & $\Re_{T}$     &17.66&17.37&16.86&\\
   &   & $\Re_{R}$     &20.33&19.99&19.32&\\
\hline
   & Fig. 6(d)& $\Lambda$ &$10^{-22}$&$10^{-21}$&$10^{-20}$&\\
   & HD & $\Delta x$ &35.16&34.61&28.45&\\
   & VD & $\Delta y$ &39.55&38.93&31.99&\\
   & & $\Re_{D}$       &56.68&55.78&45.86&\\
   &   & $\Re_{T}$     &17.65&17.38&14.26&\\
   &   & $\Re_{R}$     &20.29&19.97&16.41&\\
\hline
\end{tabular}
\end{table*}

\begin{table*}[htbp]
\centering \caption{Characteristics of the black hole
shadows in Fig. 7. \label{tab4}}
\begin{tabular}{ccccccccc}
\hline
   & Fig. 7(a)& $b_{c}$ &0&0.05&0.1&\\
   & HD & $\Delta x$    &38.61&40.69&43.01&\\
   & VD & $\Delta y$    &39.22&41.30&43.61&\\
   &    & $\Re_{D}$     &20.17&21.19&22.34&\\
   &   & $\Re_{T}$      &19.05&20.09&21.24&\\
   &   & $\Re_{R}$      &19.79&20.83&21.99&\\
\hline
   & Fig. 7(b)& $\alpha_{q}$ &0&0.1&0.15&\\
   & HD & $\Delta x$  &39.00&38.25&36.81&\\
   & VD & $\Delta y$  &39.62&38.79&37.28&\\
   & & $\Re_{D}$       &20.38&19.86&19.06&\\
   &   & $\Re_{T}$     &19.25&18.89&18.19&\\
   &   & $\Re_{R}$     &20.00&19.56&18.79&\\
\hline
   & Fig. 7(c)& $\omega_{q}$ &-0.35&-0.36&-0.37&\\
   & HD & $\Delta x$ &39.01&38.38&37.08&\\
   & VD & $\Delta y$  &39.63&38.99&37.67&\\
   & & $\Re_{D}$  &20.37&20.04&19.36&\\
   &   & $\Re_{T}$&19.25&18.93&18.30&\\
   &   & $\Re_{R}$&20.00&19.67&19.01&\\
\hline
   & Fig. 7(d)& $\Lambda$ &$10^{-22}$&$10^{-21}$&$10^{-20}$&\\
   & HD & $\Delta x$ &38.94&38.32&31.50&\\
   & VD & $\Delta y$ &39.55&38.93&31.99&\\
   & & $\Re_{D}$       &20.33&20.01&16.45&\\
   &   & $\Re_{T}$     &19.21&18.91&15.55&\\
   &   & $\Re_{R}$     &19.96&19.64&16.15&\\
\hline
\end{tabular}
\end{table*}

\begin{table*}[htbp]
\centering \caption{Characteristics of the black hole
shadows in Fig. 8.\label{tab5}}
\begin{tabular}{ccccccccc}
\hline
   & Fig. 8(a)& $b_{c}$ &0&0.05&0.1&\\
   & HD & $\Delta x$    &39.22&41.28&43.59&\\
   & VD & $\Delta y$    &39.20&41.30&43.61&\\
   &    & $\Re_{D}$     &19.62&20.66&21.82&\\
   &   & $\Re_{T}$      &19.58&20.63&21.78&\\
   &   & $\Re_{R}$      &19.62&20.66&21.82&\\
\hline
   & Fig. 8(b)& $\alpha_{q}$ &0&0.1&0.15&\\
   & HD & $\Delta x$  &39.60&38.77&37.26&\\
   & VD & $\Delta y$  &39.62&38.79&37.28&\\
   & & $\Re_{D}$       &19.82&19.40&18.65&\\
   &   & $\Re_{T}$     &19.79&19.37&18.62&\\
   &   & $\Re_{R}$     &19.82&19.40&18.65&\\
\hline
   & Fig. 8(c)& $\omega_{q}$ &-0.35&-0.36&-0.37&\\
   & HD & $\Delta x$ &39.60&38.97&37.65&\\
   & VD & $\Delta y$  &39.62&38.99&37.67&\\
   & & $\Re_{D}$  &19.82&19.51&18.84&\\
   &   & $\Re_{T}$&19.79&19.47&18.81&\\
   &   & $\Re_{R}$&19.82&19.51&18.84&\\
\hline
   & Fig. 8(d)& $\Lambda$ &$10^{-22}$&$10^{-21}$&$10^{-20}$&\\
   & HD & $\Delta x$ &39.53&38.90&31.98&\\
   & VD & $\Delta y$ &39.55&38.93&32.00&\\
   & & $\Re_{D}$       &19.79&19.48&16.01&\\
   &   & $\Re_{T}$     &19.75&19.44&15.98&\\
   &   & $\Re_{R}$     &19.79&19.47&16.01&\\
\hline
\end{tabular}
\end{table*}

\begin{table*}[htbp]
\centering \caption{Characteristics of the black hole
shadows in Fig. 9.  \label{tab6}}
\begin{tabular}{ccccccccc}
\hline
   & Fig. 9(a)& $a$ &0.5&0.8&0.94&\\
   & HD & $\Delta x$    &38.91&37.84&36.99&\\
   & VD & $\Delta y$    &38.97&38.04&37.37&\\
   &    & $\Re_{D}$     &19.52&19.17&19.03&\\
   &   & $\Re_{T}$      &19.43&18.82&18.33&\\
   &   & $\Re_{R}$      &19.51&19.09&18.80&\\
\hline
   & Fig. 9(b)& $a$ &0.5&0.8&0.94&\\
   & HD & $\Delta x$  &38.91&37.84&36.99&\\
   & VD & $\Delta y$  &38.97&38.04&37.37&\\
   & & $\Re_{D}$       &19.52&19.17&19.03&\\
   &   & $\Re_{T}$     &19.43&18.83&18.33&\\
   &   & $\Re_{R}$     &19.51&19.09&18.80&\\
\hline
   & Fig. 9(c)& $a$ &0.5&0.8&0.94&\\
   & HD & $\Delta x$ &38.91&37.84&36.99&\\
   & VD & $\Delta y$  &38.97&38.04&37.37&\\
   & & $\Re_{D}$  &19.52&19.17&19.03&\\
   &   & $\Re_{T}$&19.43&18.83&18.33&\\
   &   & $\Re_{R}$&19.51&19.09&18.80&\\
\hline
   & Fig. 9(d)& $a$ &0.5&0.8&0.94&\\
   & HD & $\Delta x$ &39.01&37.88&36.97&\\
   & VD & $\Delta y$ &39.09&38.09&37.37&\\
   & & $\Re_{D}$       &19.57&19.20&19.07&\\
   &   & $\Re_{T}$     &19.49&18.83&18.31&\\
   &   & $\Re_{R}$     &19.56&19.12&18.80&\\
\hline
\end{tabular}
\end{table*}

\begin{figure*}[htbp]
\center{
\includegraphics[scale=0.3]{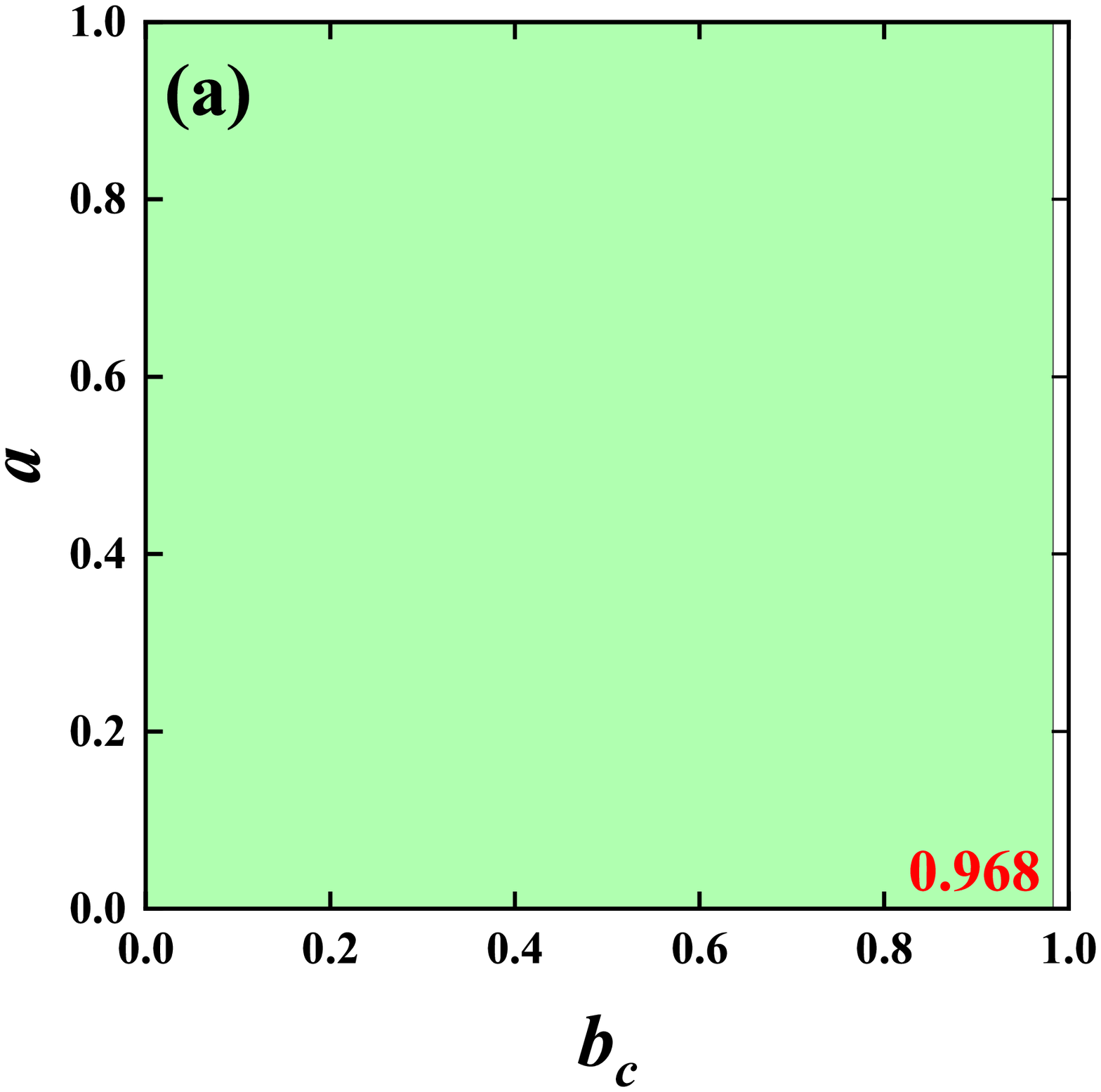}
\includegraphics[scale=0.3]{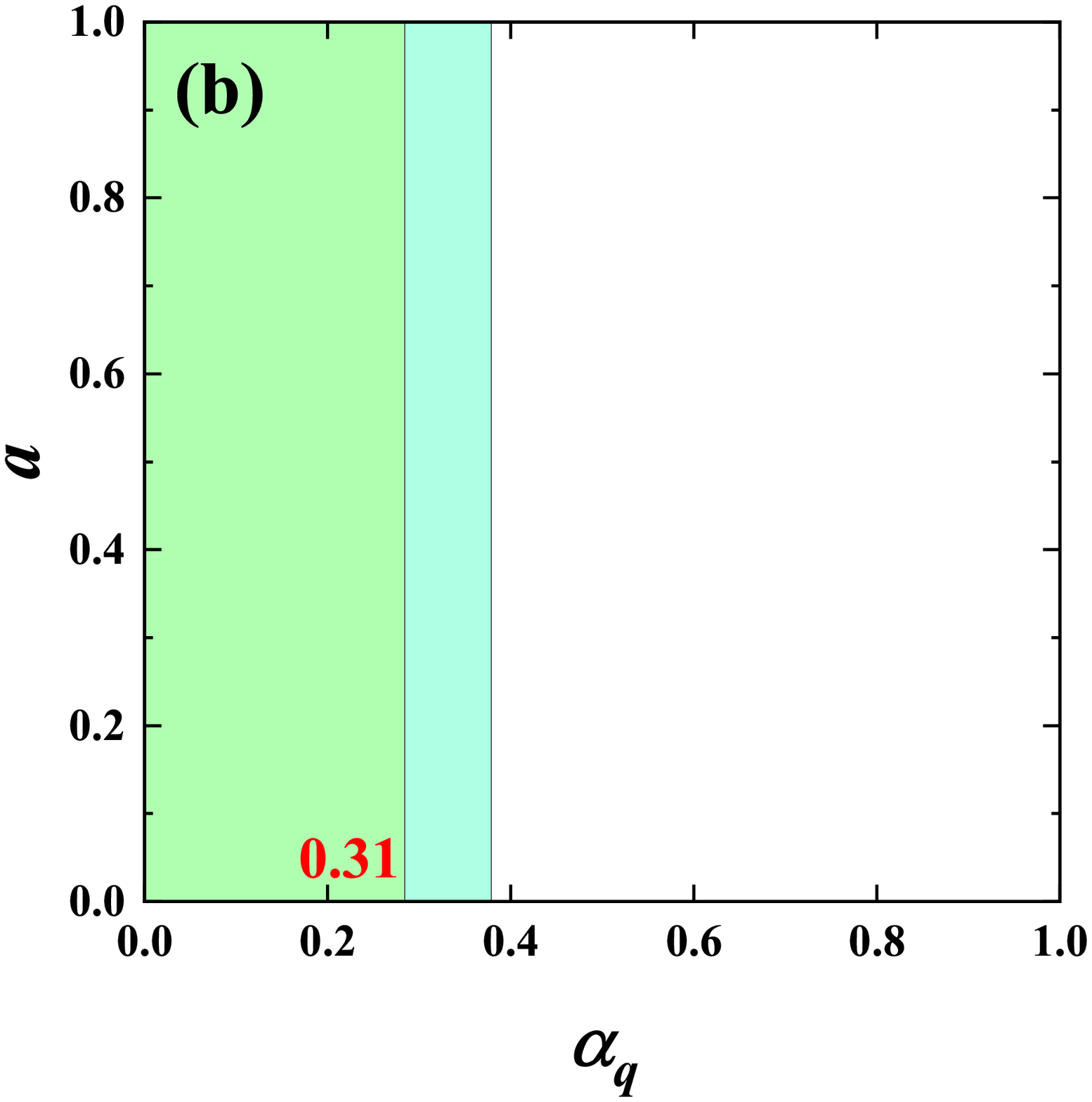}
\includegraphics[scale=0.3]{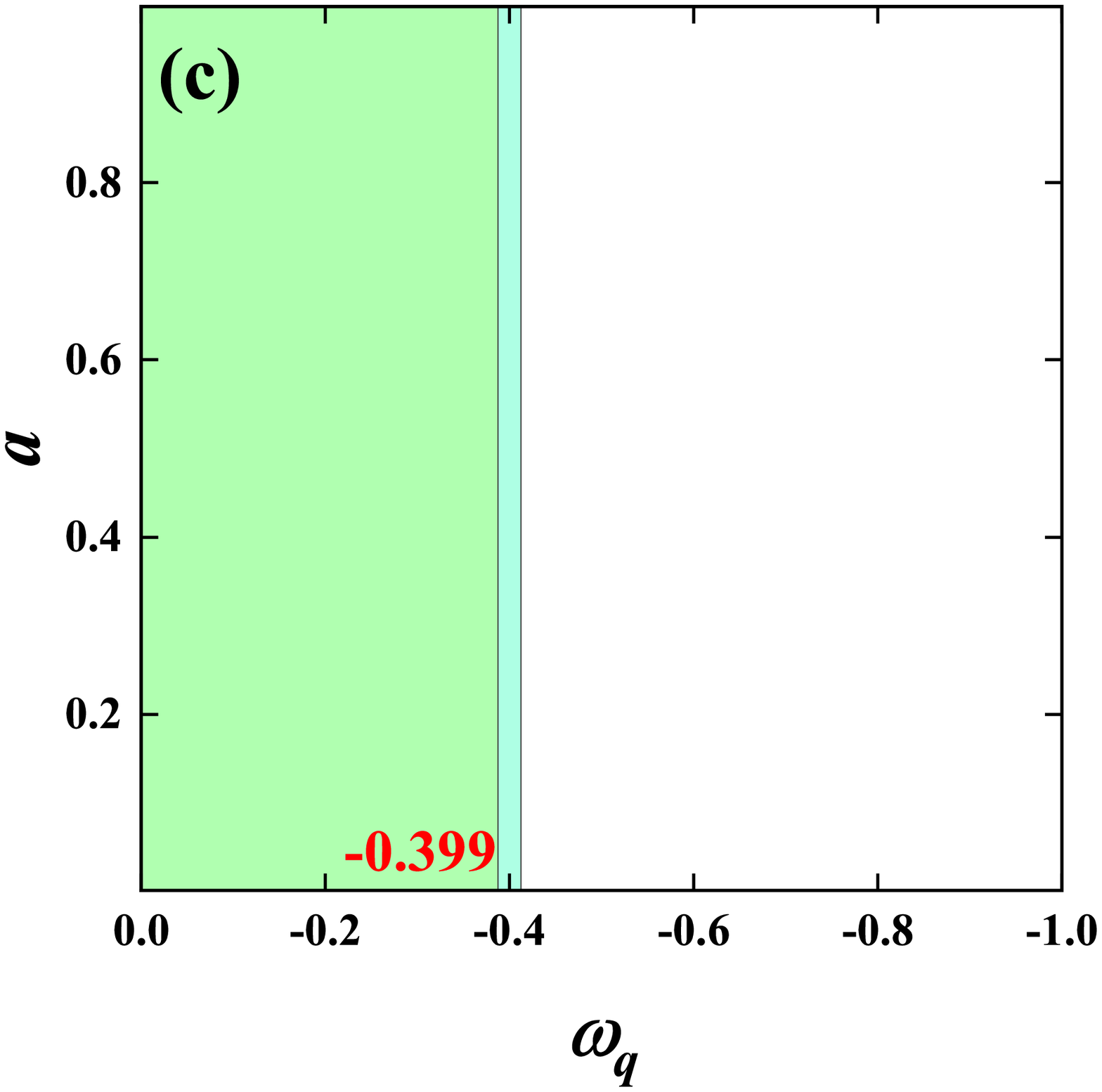}
\includegraphics[scale=0.3]{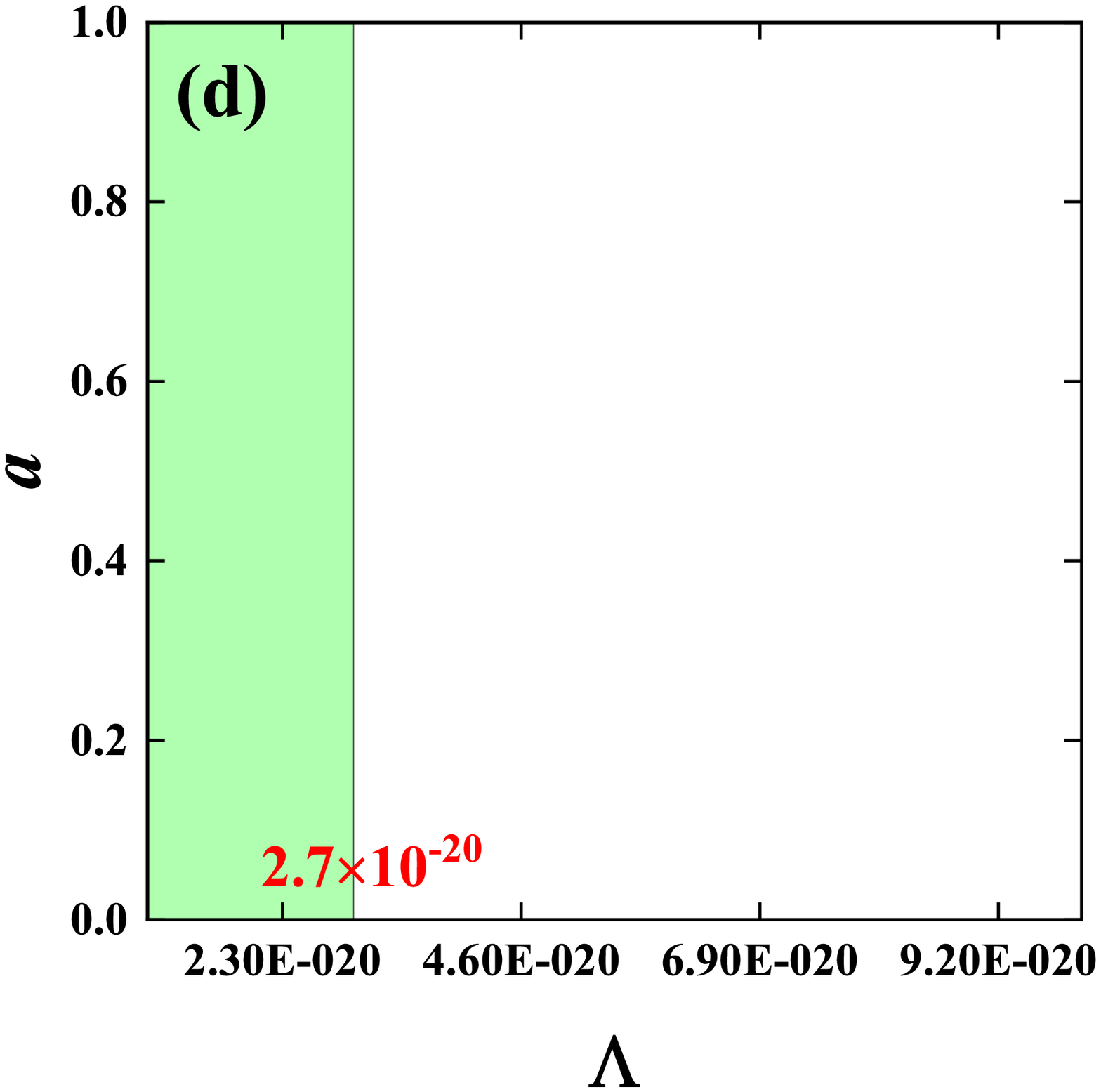}
\caption{(a) The domain of outer communication on the parameters $a$ and $b_c$ with $Q=0.2$, $\alpha_{q}=0.01$, $\omega_{q}=-0.35$ and $\Lambda=1.02\times10^{-26}$.
(b) The domain on the parameters $a$ and $\alpha_q$ with $Q=0.2$, $b_{c}=0.01$, $\omega_{q}=-0.35$ and $\Lambda=1.02\times10^{-26}$.
(c) The domain on the parameters $a$ and $\omega_q$ with $Q=0.2$ and $b_{c}=0.01$, $\alpha_{q}=0.01$ and $\Lambda=1.02\times10^{-26}$.
(d) The domain on the parameters $a$ and $\Lambda$ with
$Q=0.2$, $b_{c}=0.01$, $\alpha_{q}=0.01$ and $\omega_{q}=-0.35$. Assume that the distance from the observer to the black hole is $r_0= 10^{10}M$.
In the green regions for $r_c>r_0$, light rays can reach a rest observer's eyes; but they cannot in  the blue regions for $r_c<r_0$.
The white regions represent $r_c$ as imaginary numbers.
}\label{Fig1}}
\end{figure*}

\begin{figure*}[htbp]
\center{
\includegraphics[scale=0.2]{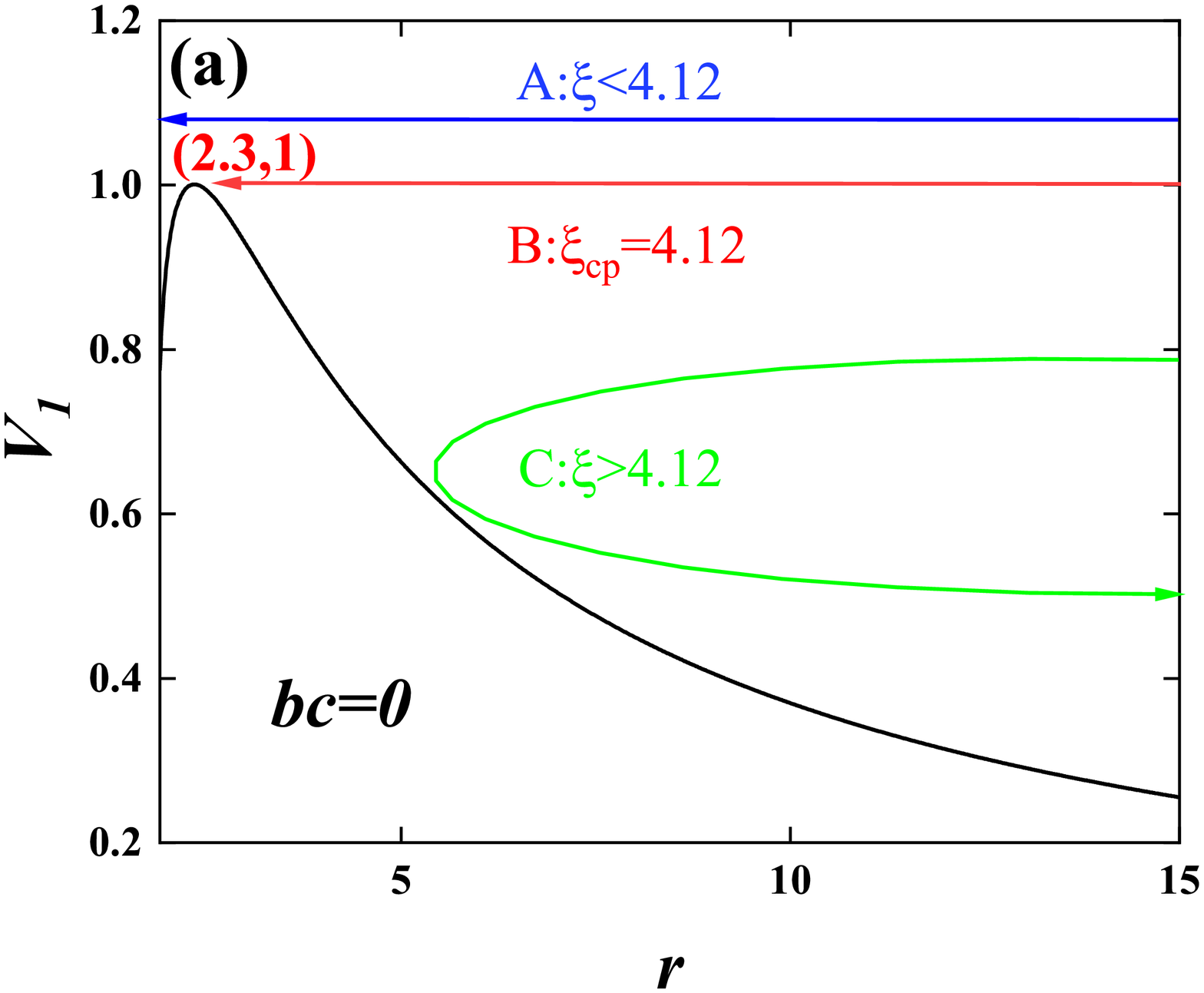}
\includegraphics[scale=0.2]{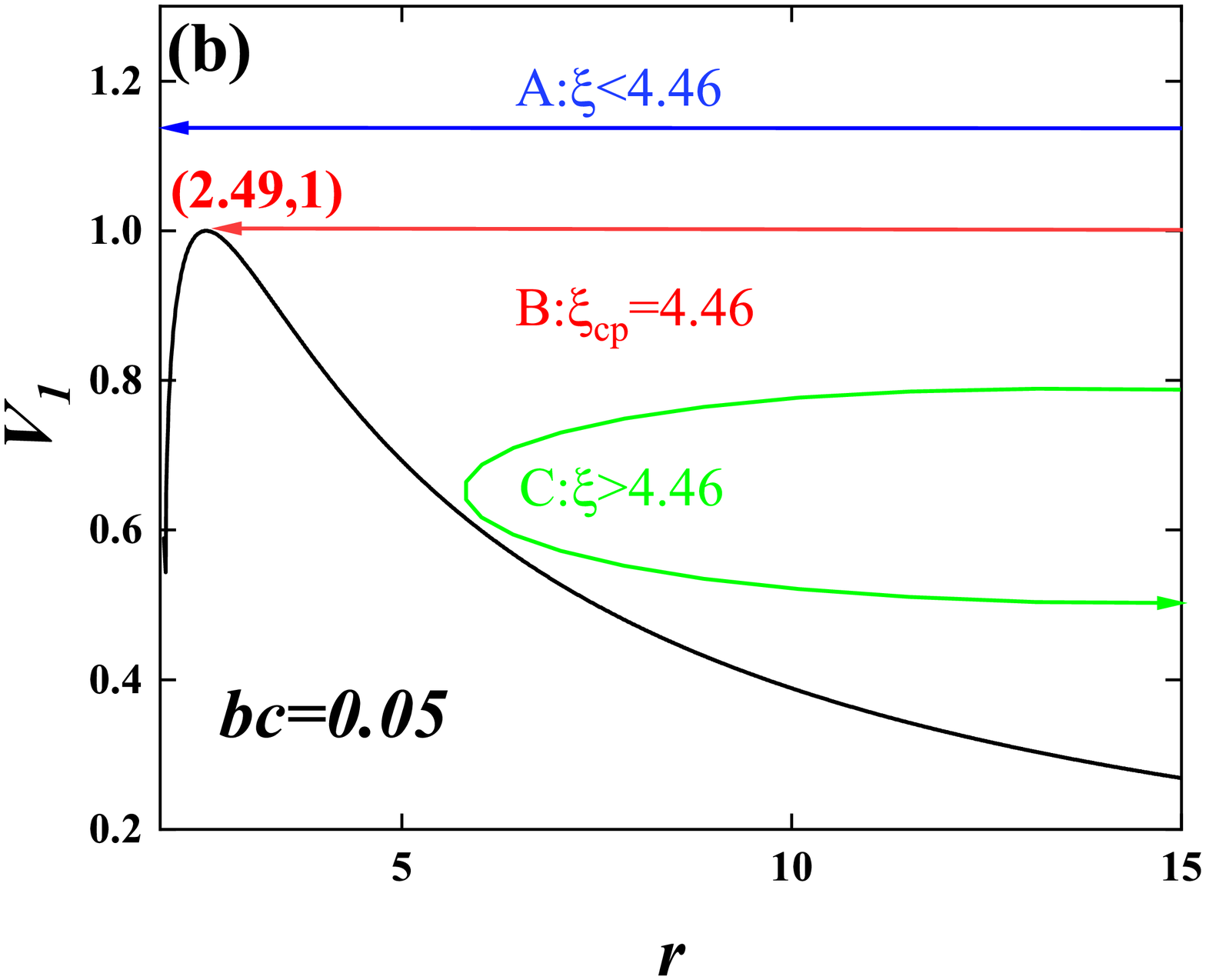}
\includegraphics[scale=0.2]{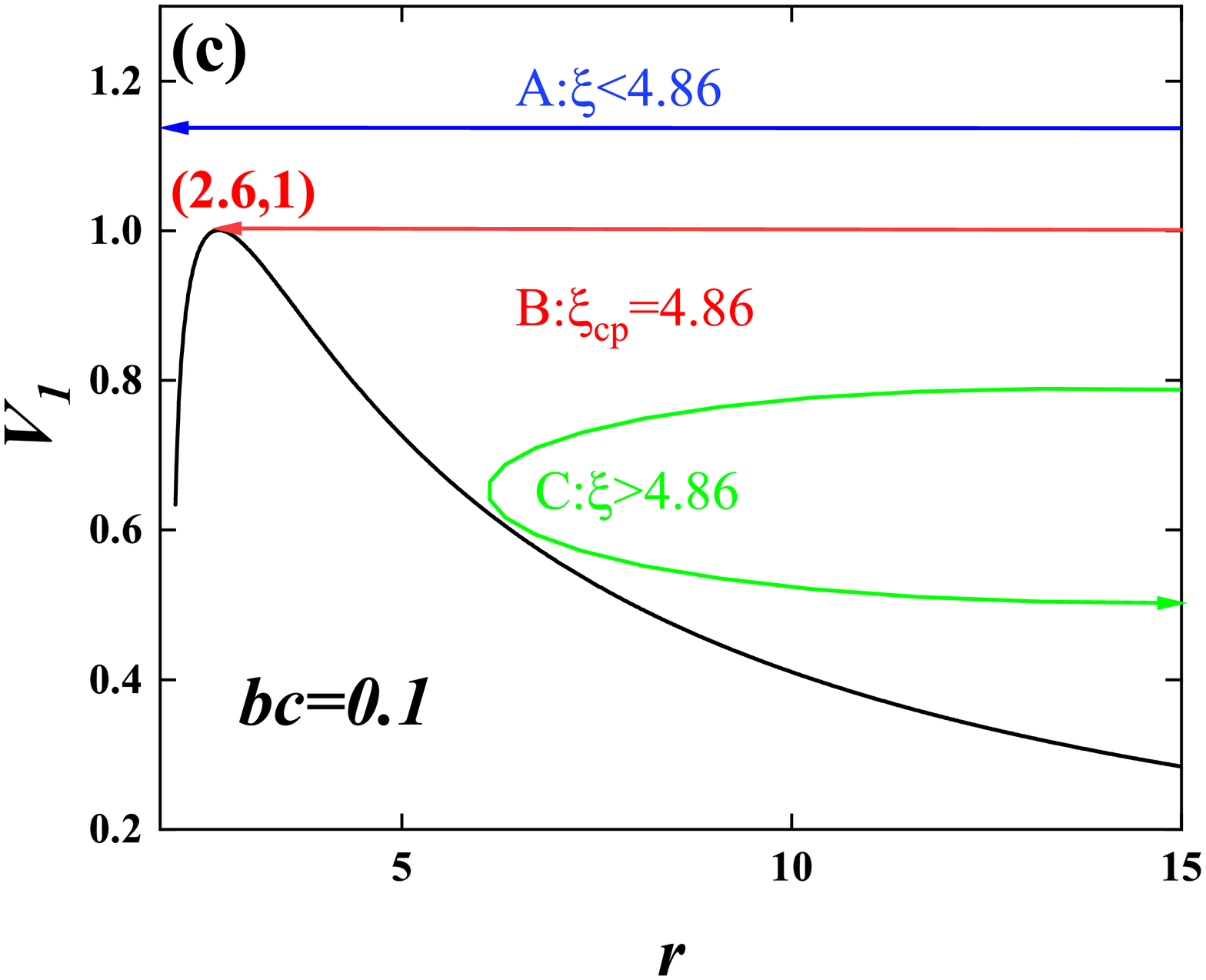}
\includegraphics[scale=0.2]{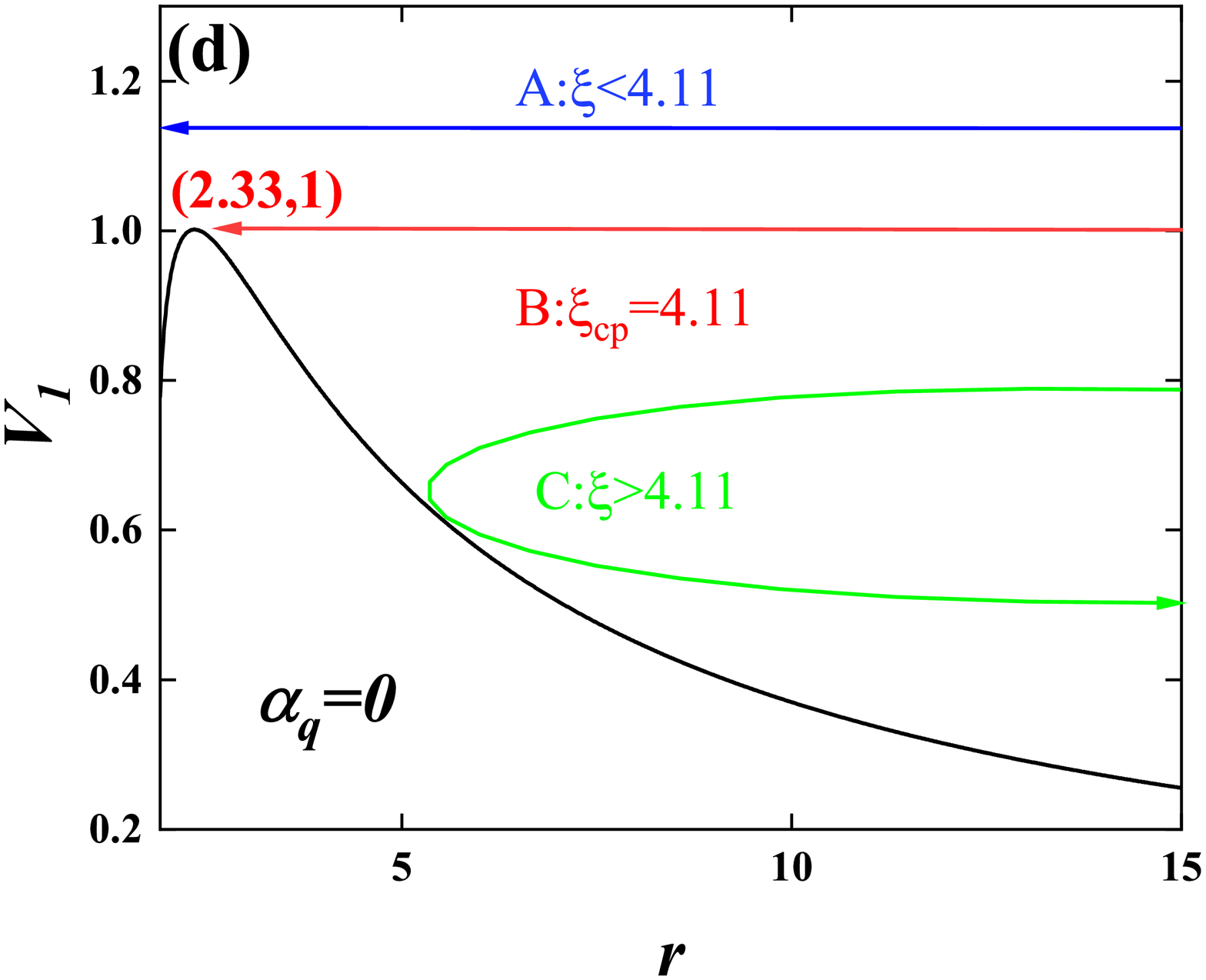}
\includegraphics[scale=0.2]{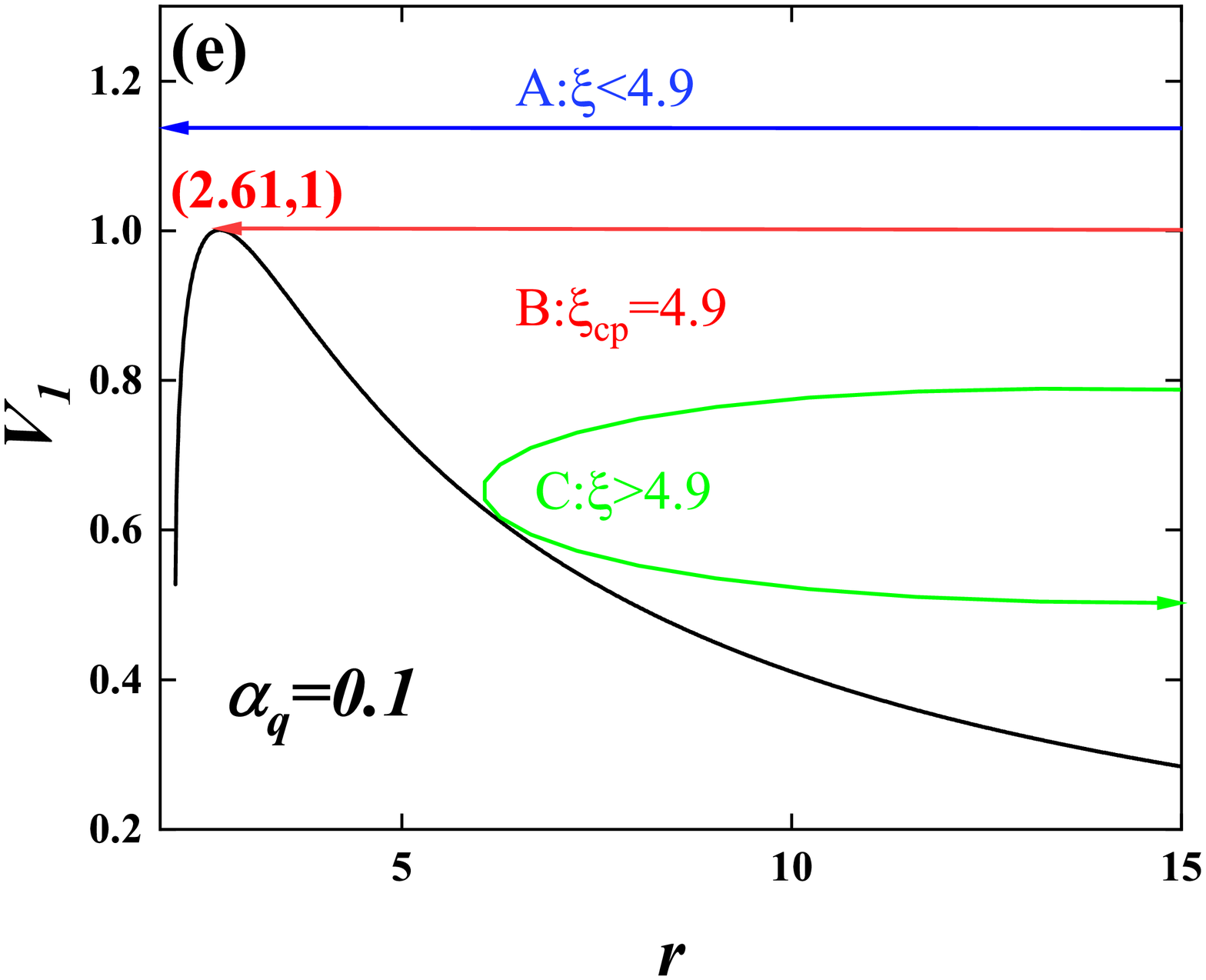}
\includegraphics[scale=0.2]{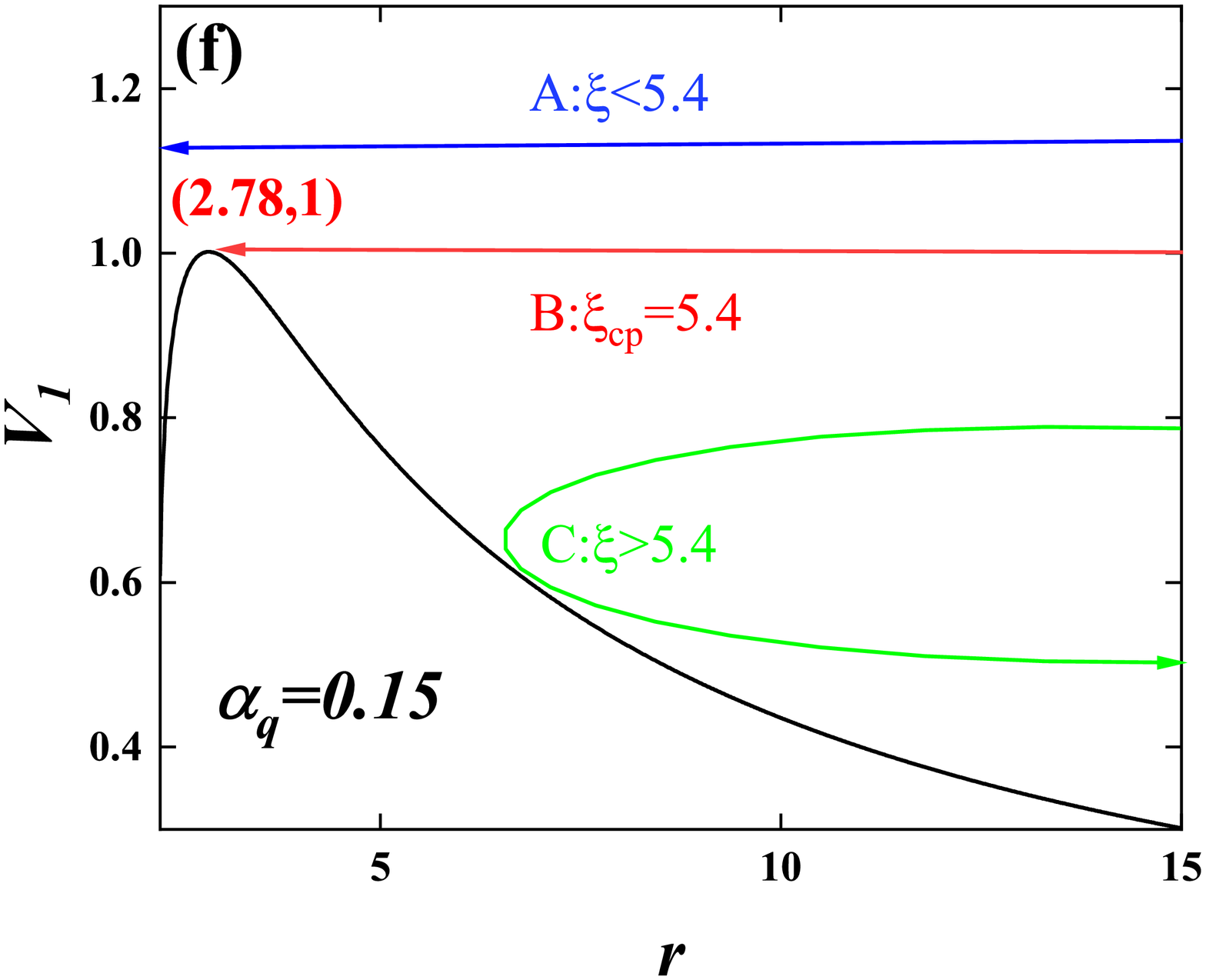}
\includegraphics[scale=0.2]{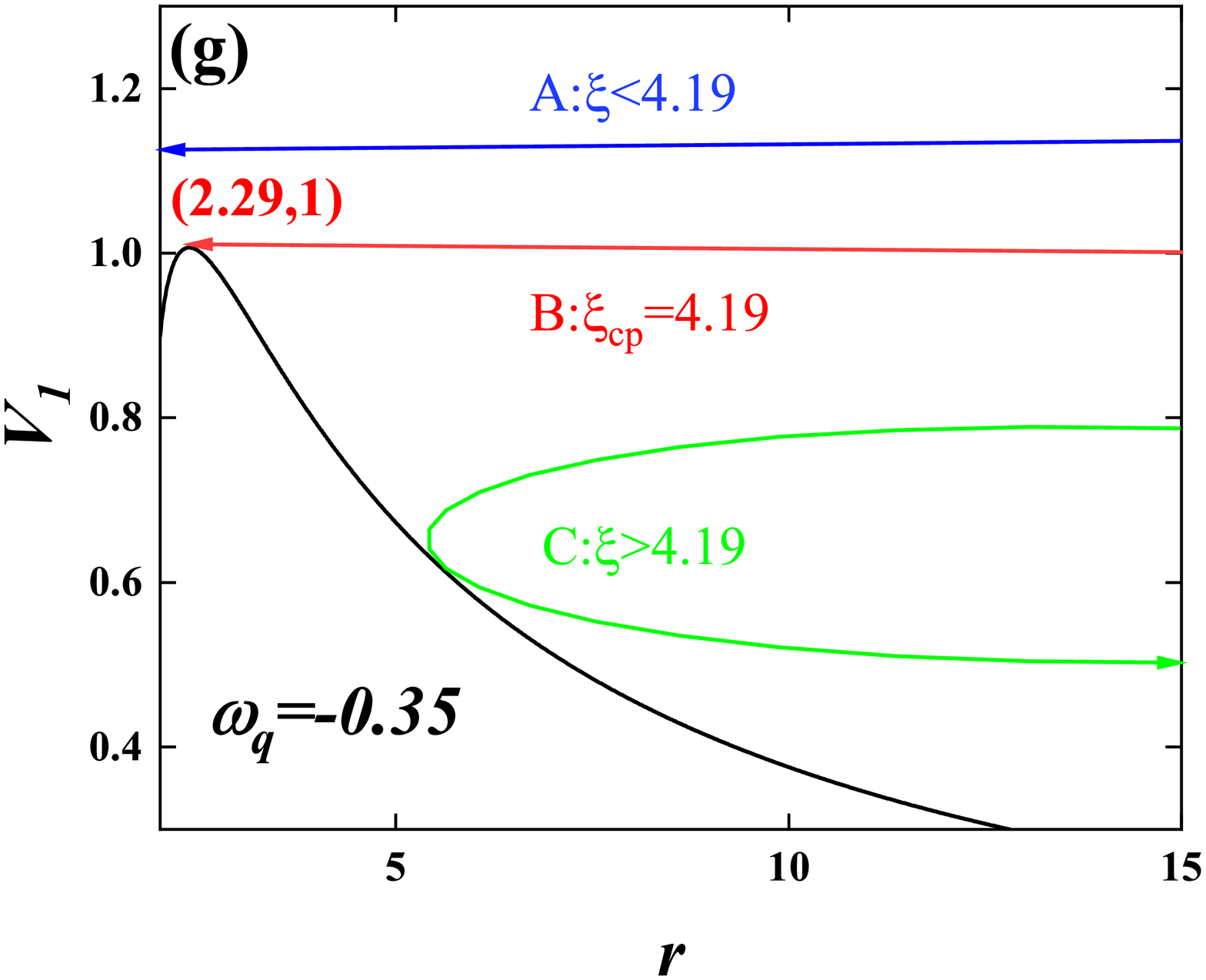}
\includegraphics[scale=0.2]{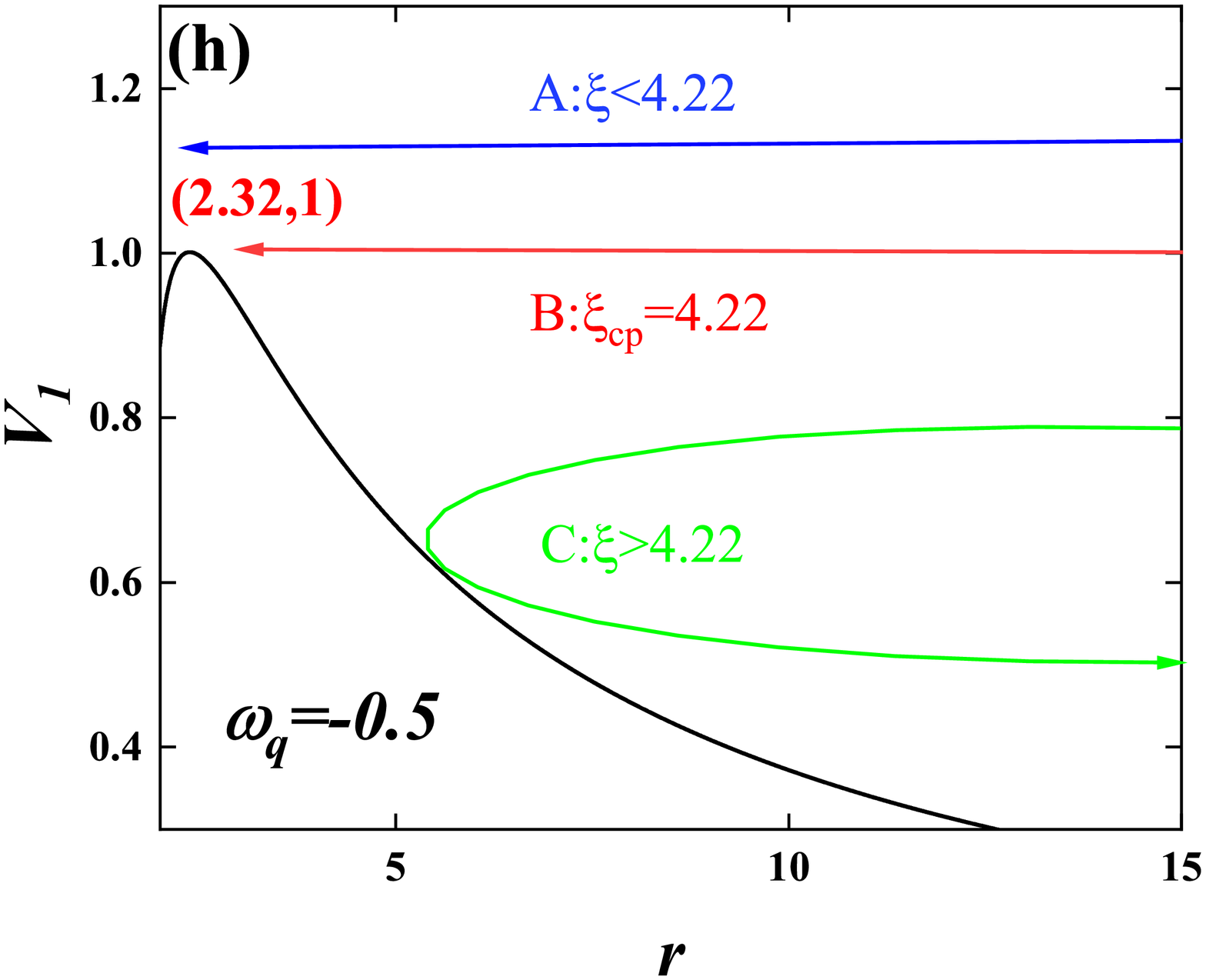}
\includegraphics[scale=0.2]{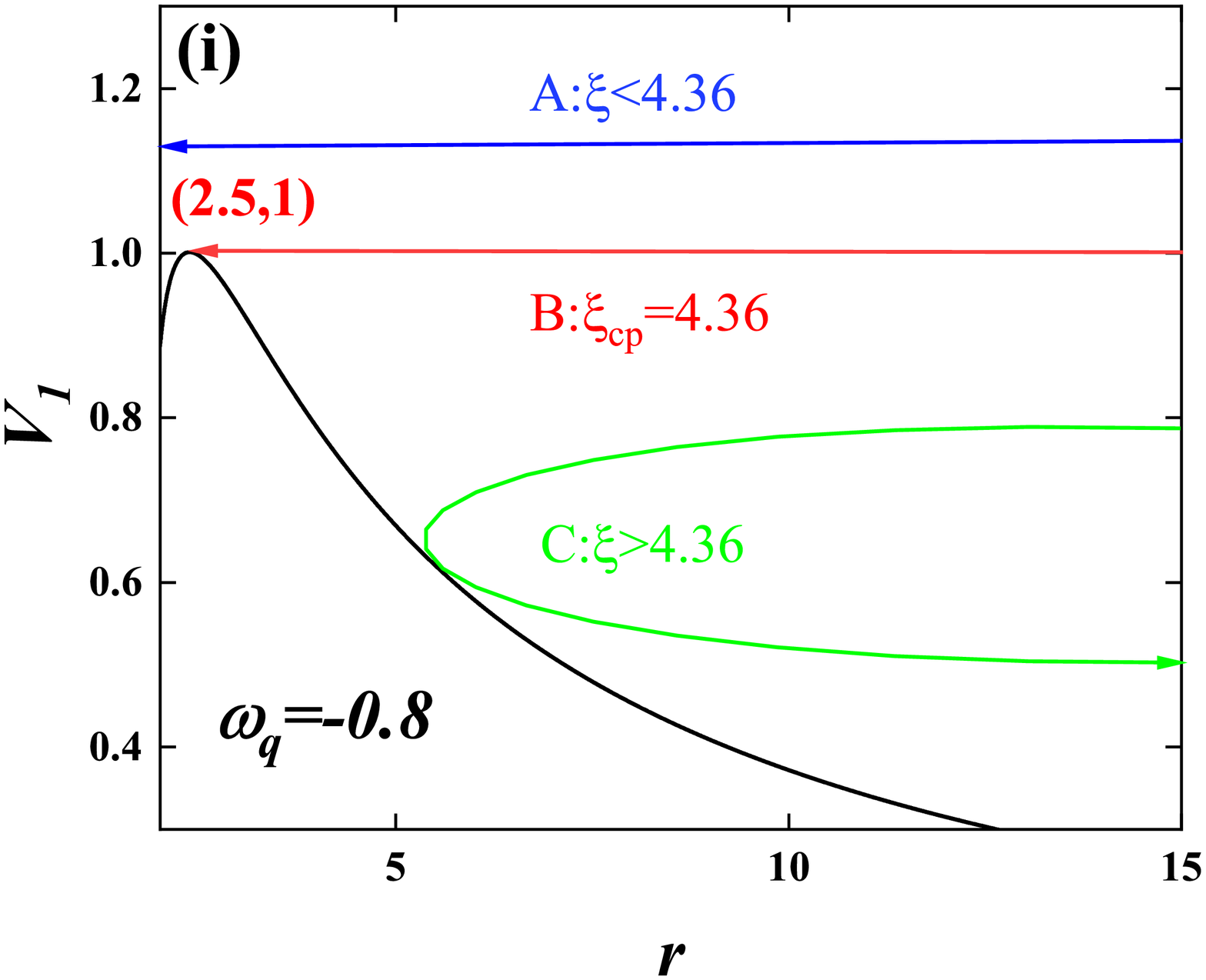}
\caption{ Effective potentials for the motion of photons on the
equatorial plane $\vartheta=\pi/2$ around the
black holes  in several combinations of the parameters. The parameters in these panels are $a=0.5$,
$Q=0.2$, $\Lambda=1.2\times10^{-26}$ and $\theta_0=\pi/2$. (a)-(c): $\alpha_q=0.01$ and $\omega_q=-0.35$, but three different values
are given to $b_c$. (d)-(f): $b_c=0.01$ and $\omega_q=-0.35$, but three different values
are given to $\alpha_q$. (g)-(i): $b_c=0.01$ and $\alpha_q=0.1$, but three different values
are given to $\omega_q$. The maximum values correspond to unstable circular
photon orbits. The circular orbital radius $r_{cp}$ increases with each of the parameters
$b_c$, $\alpha_{q}$ and  $|\omega_{q}|$ increasing.
The effects of the parameters on the effective
potentials of spherical orbits are the same as those on the circular orbits.
The cosmological constant has little influence on them.
 } \label{Fig2}}
\end{figure*}

\begin{figure*}[htbp]
\center{
\includegraphics[scale=0.4]{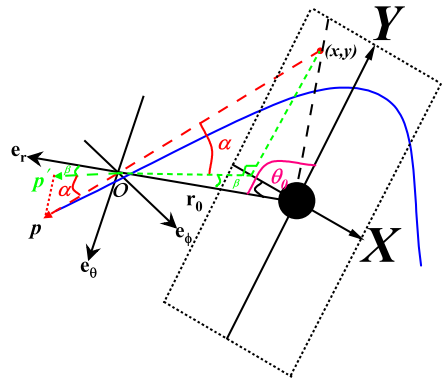}
\caption{Projection of the photon momentum $\vec{p}$ in the
observer's frame and the solid angles $(\alpha,\beta)$ [46-48].}
\label{Fig3}}
\end{figure*}

\begin{figure*}[htbp]
\center{
\includegraphics[scale=0.25]{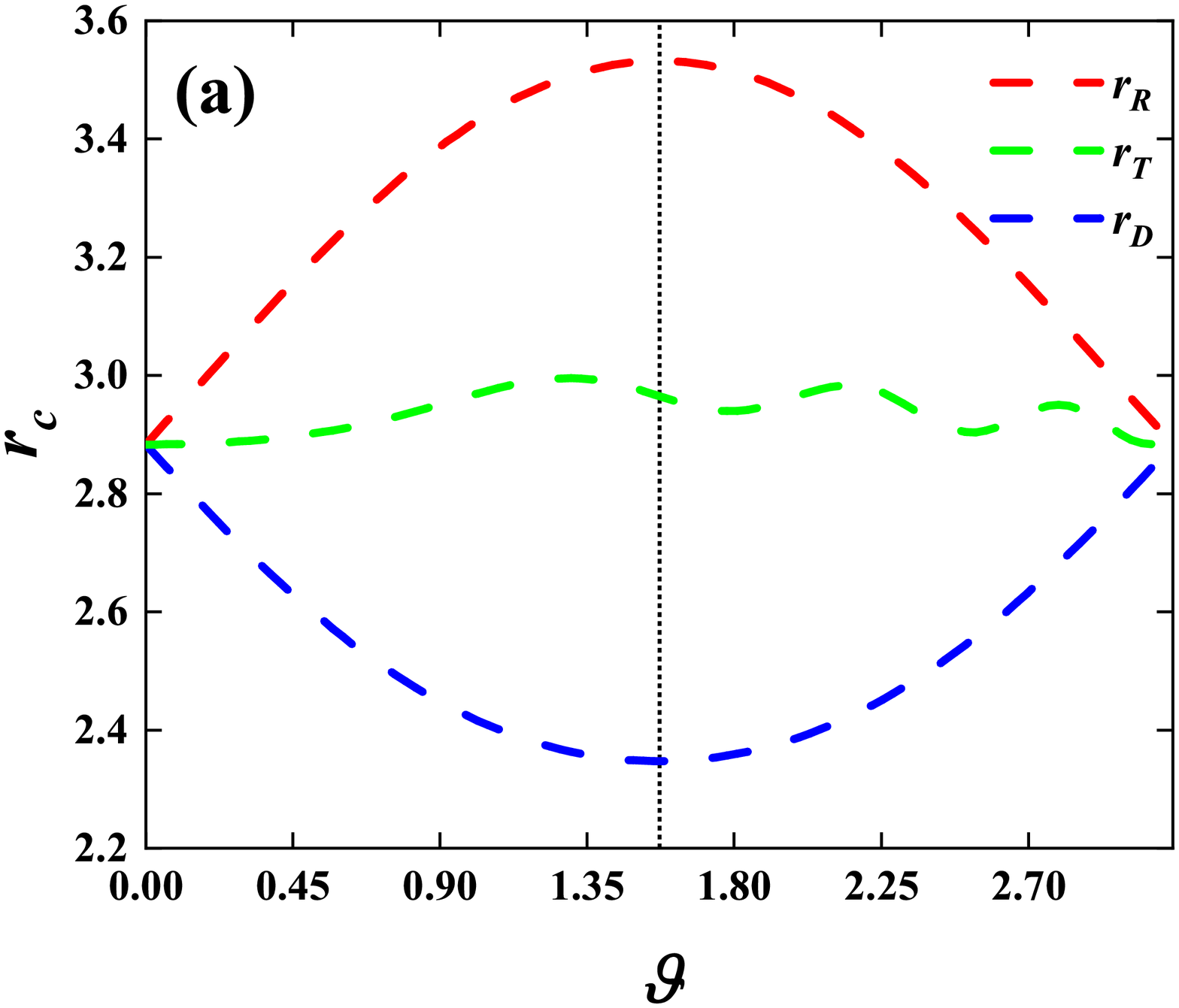}
\includegraphics[scale=0.25]{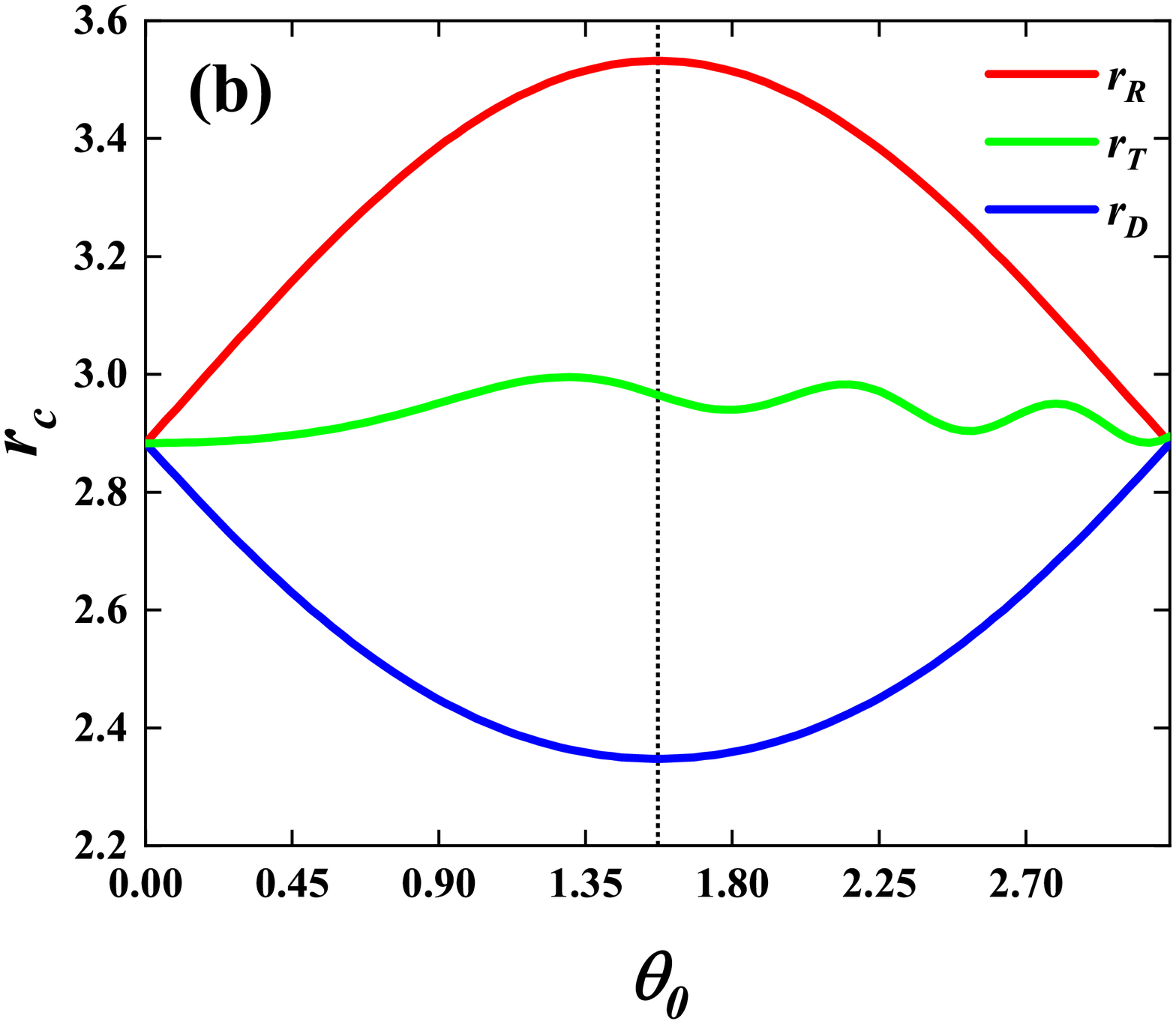}
\includegraphics[scale=0.25]{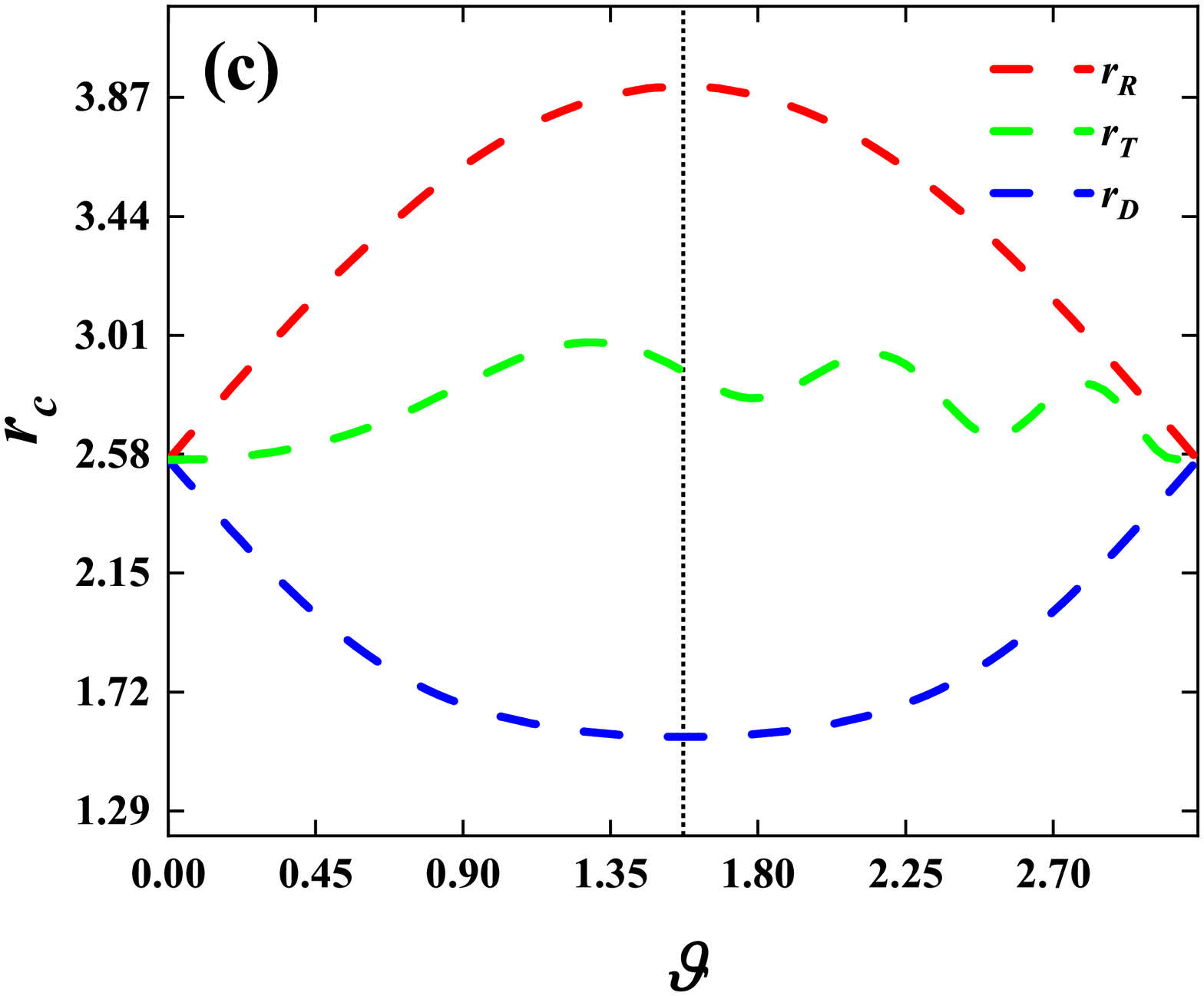}
\includegraphics[scale=0.25]{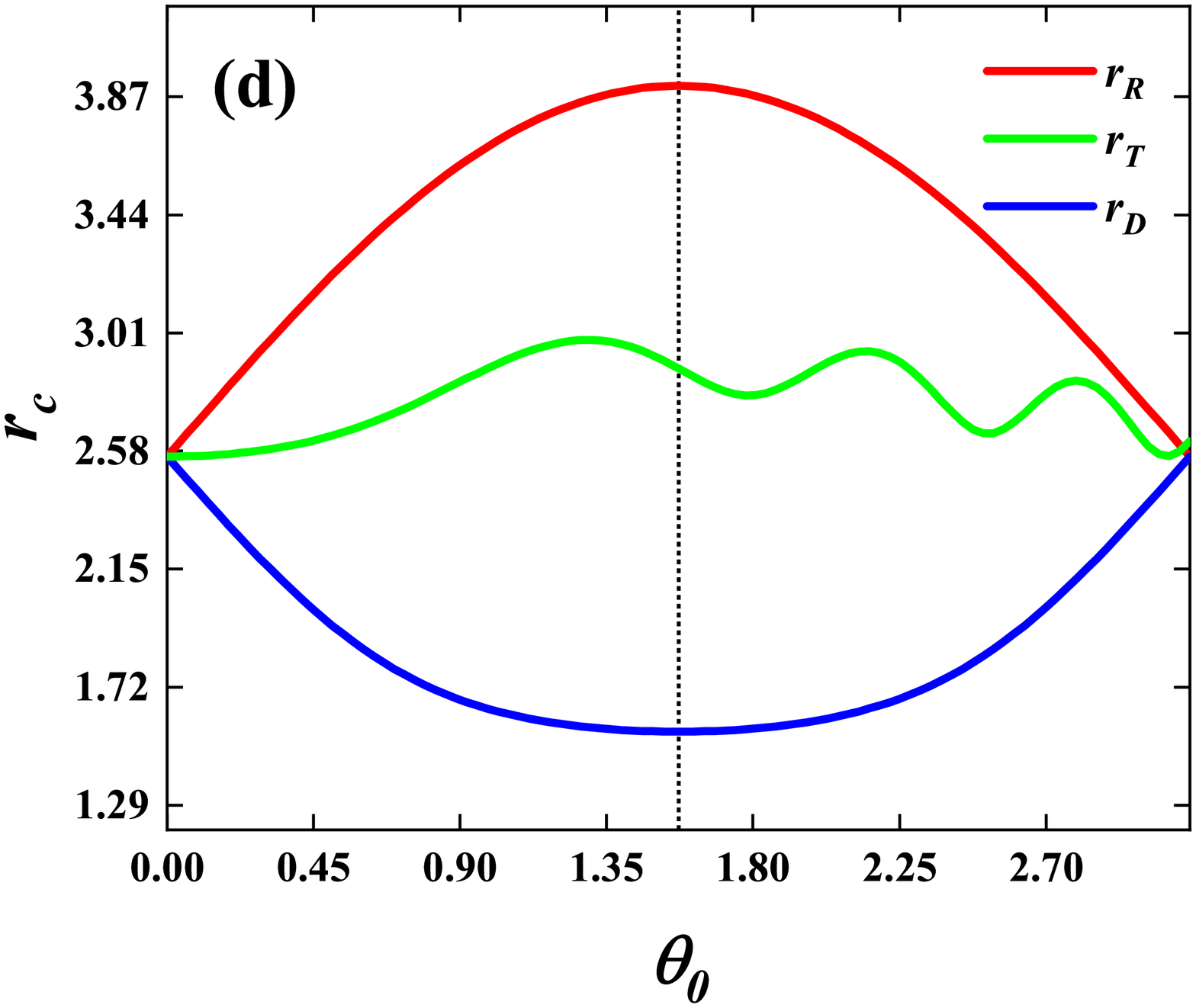}
\includegraphics[scale=0.25]{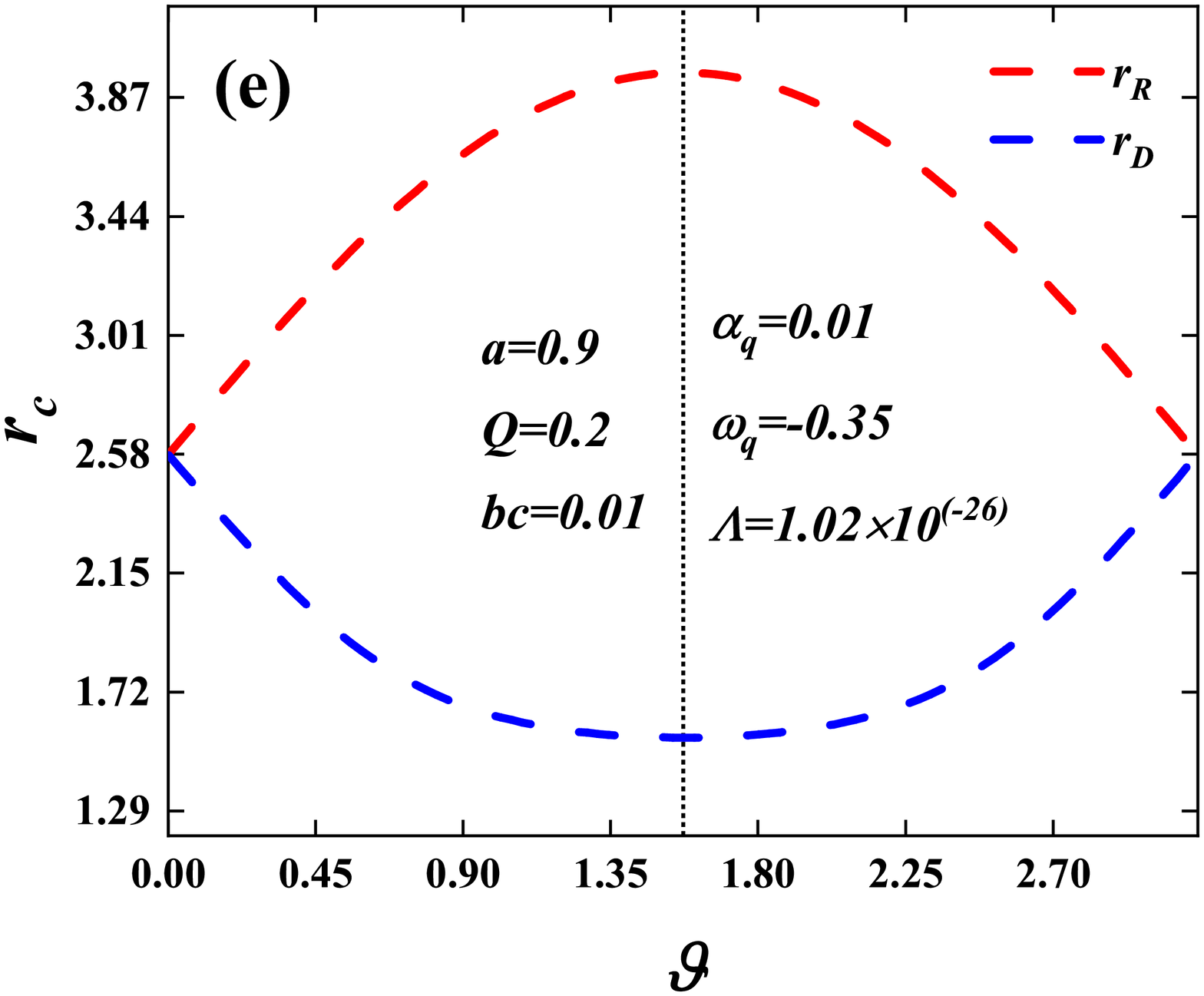}
\includegraphics[scale=0.25]{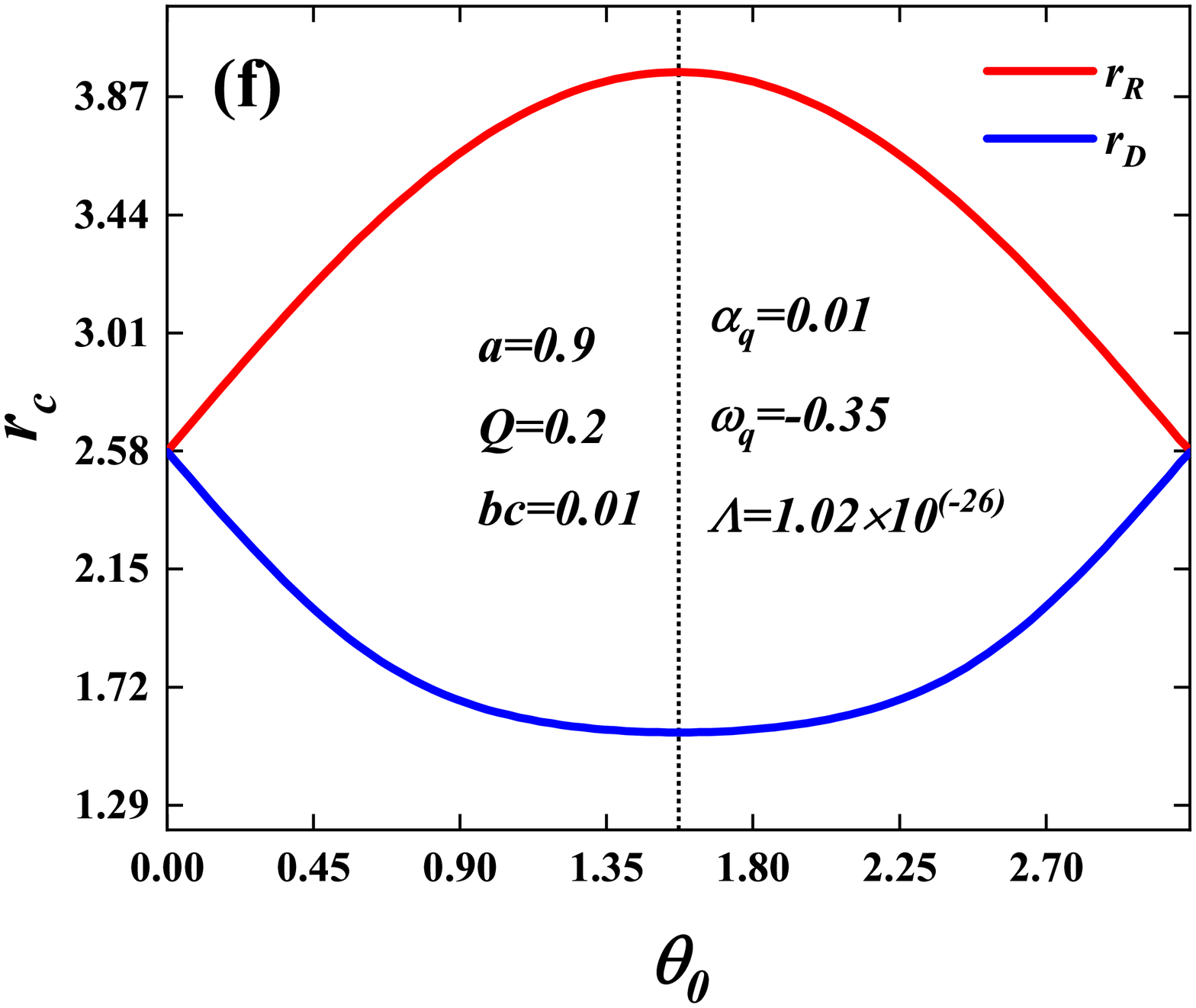}
\caption{Dependence of the critical photon orbital radii $r_D$ and $r_R$
on observational angle $\theta_{0}$. The radii are
obtained from Method 1 in the left-hand side, and  Method 2 in the right-hand side.
(a) and (b): For the Kerr black hole with $a=0.5$, circular photon orbits
stay in all planes $\vartheta=\theta_0$. (c) and (d): Same as (a) and (b) but $a=0.9$.
The green curves in these four panels represent the circular photon orbits that determine
the characteristic points $T$ introduced in Ref. [48]. (e) and (f): The critical photon orbits
in the  KNdS spacetime.  } \label{Fig4}}
\end{figure*}

\begin{figure*}[htbp]
\center{
\includegraphics[scale=0.3]{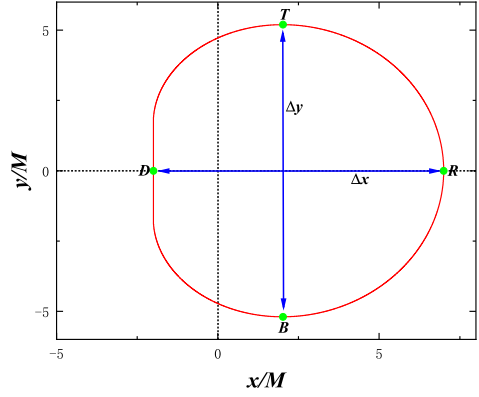}
\caption{Characteristic points of the Kerr black hole shadow.
The red circle denotes the shape of the shadow for the spin $a$
and observation angle $\theta_{0}$. The characteristic points $D$,
$R$, $T$ and $B$ respectively correspond to the left, right, top,
and bottom points of the shadow. $\Delta x$ and $\Delta y$ denote
the horizontal and vertical diameters of the shadow.}
\label{Fig5}}
\end{figure*}

\begin{figure*}[htbp]
\center{
\includegraphics[scale=0.5]{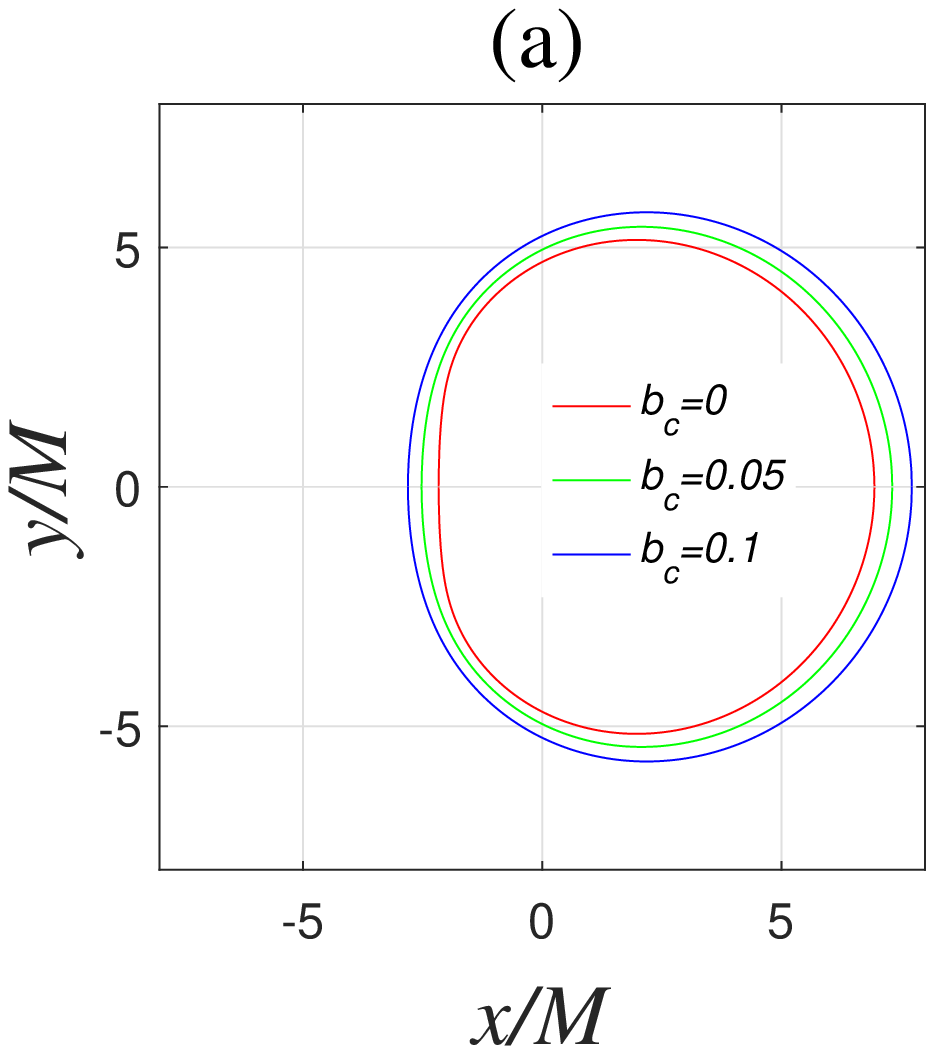}
\includegraphics[scale=0.5]{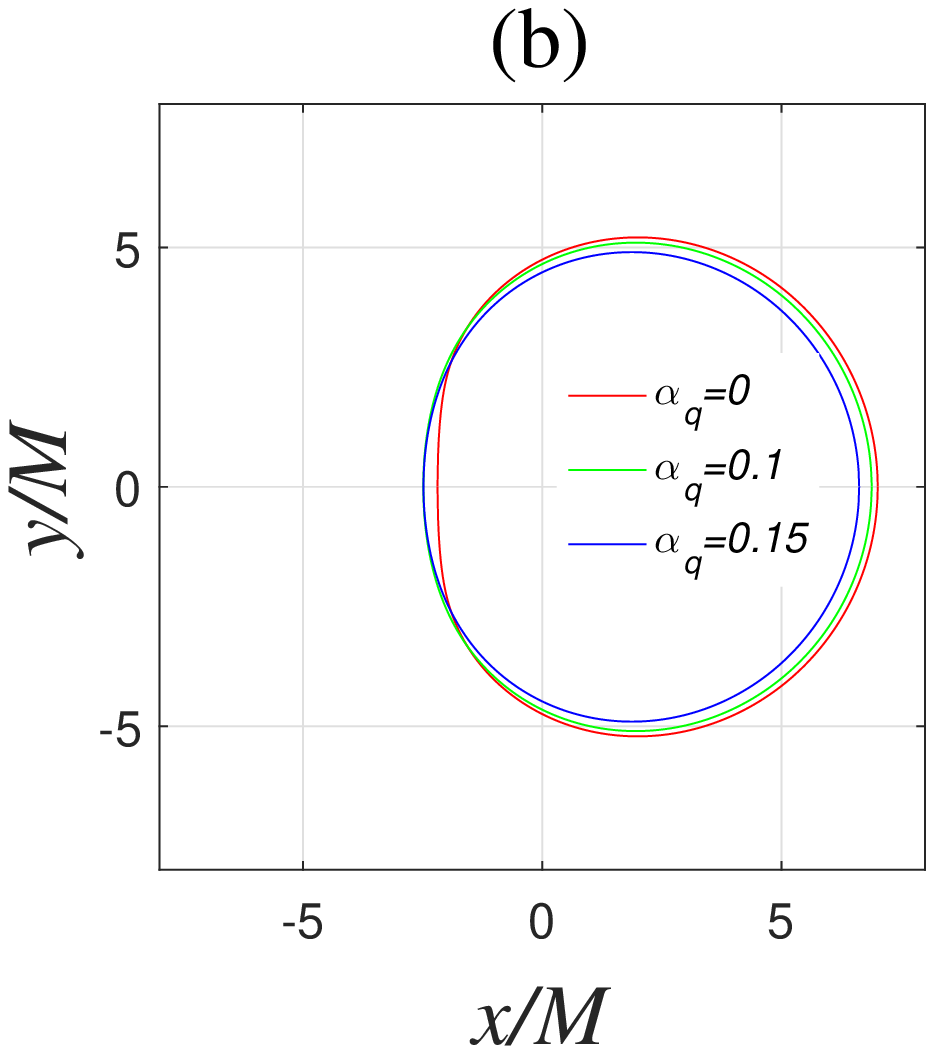}
\includegraphics[scale=0.5]{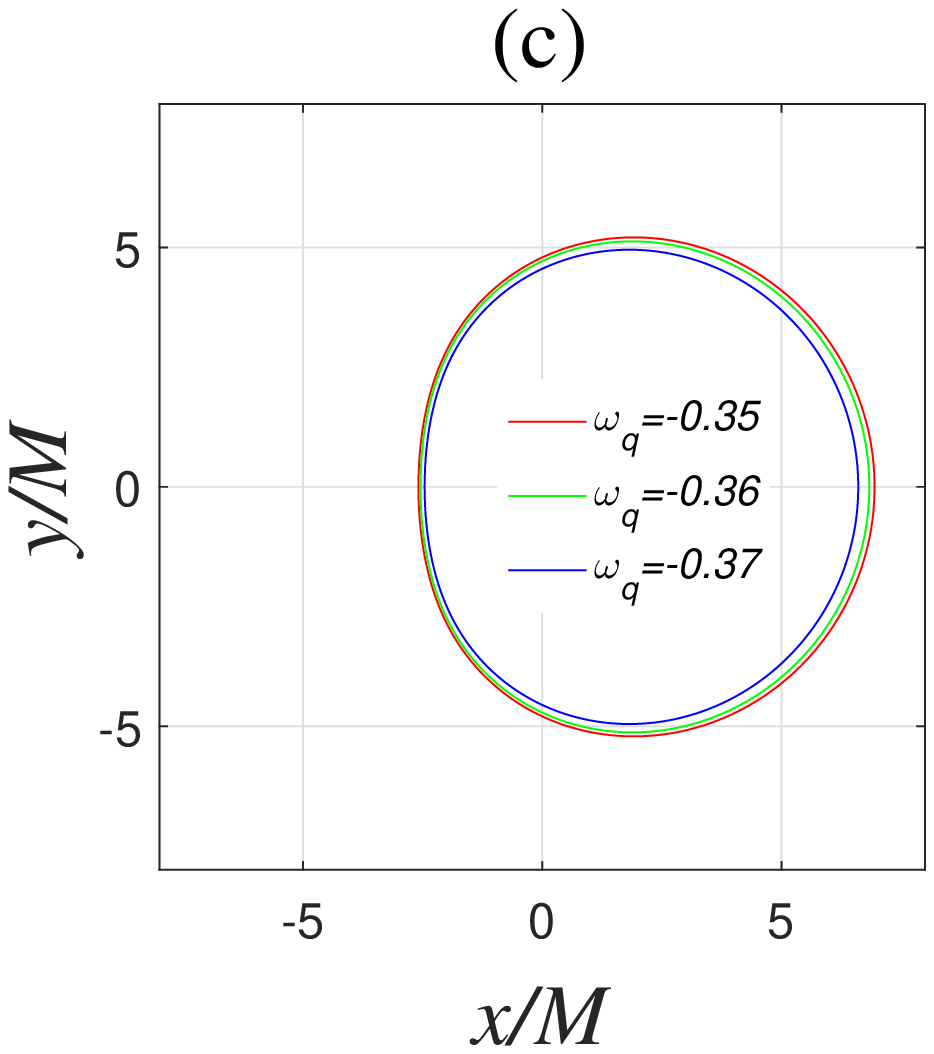}
\includegraphics[scale=0.5]{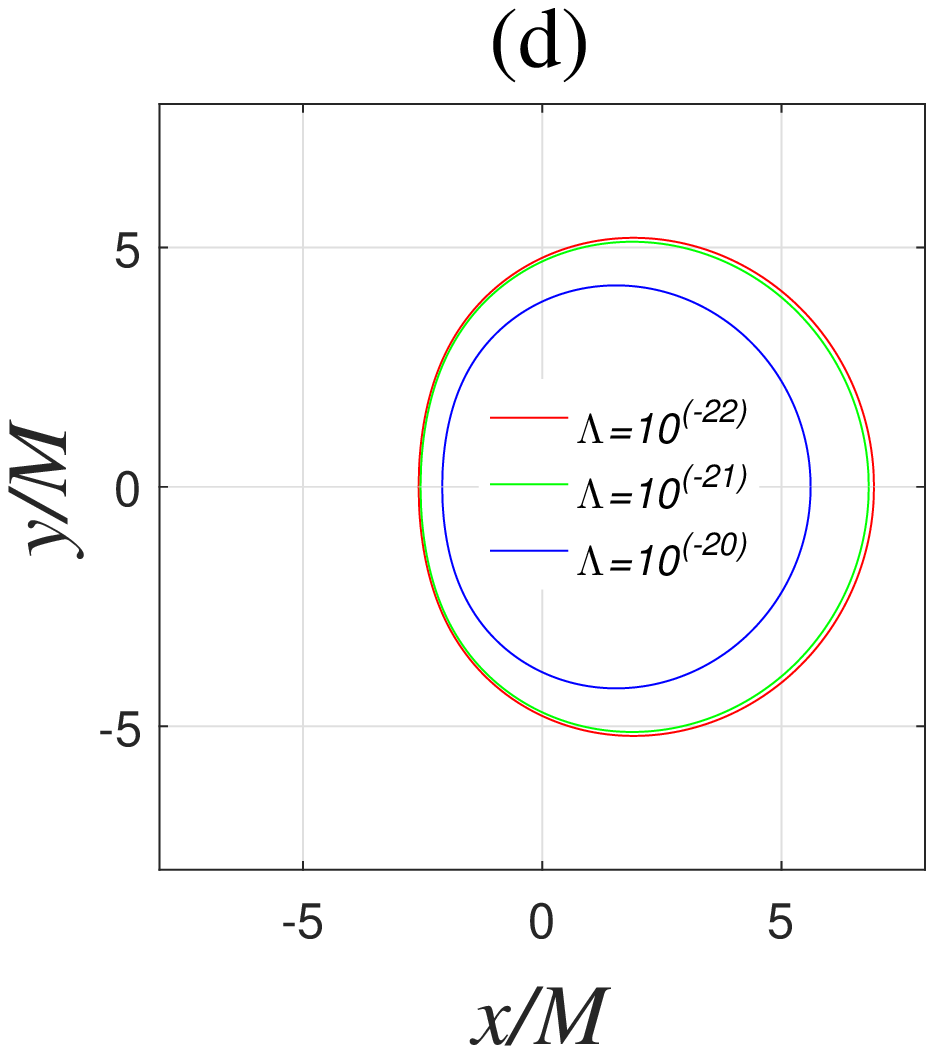}
\caption{Black hole shadows for several combinations of the parameters. The parameters in these panels are $a=0.98$, $\theta_0=90^{\circ}$, and $Q=0.2$.
(a) $\alpha_q=0.01$, $\omega_q=-0.35$ and $\Lambda=1.02\times10^{-26}$.
(b) $b_{c}=0.01$, $\omega_q=-0.35$ and $\Lambda=1.02\times10^{-26}$.
(c) $b_{c}=0.01$, $\alpha_q=0.01$ and $\Lambda=1.02\times10^{-26}$.
(d) $b_{c}=0.01$, $\alpha_q=0.01$ and $\omega_q=-0.35$.}
\label{Fig6}}
\end{figure*}

\begin{figure*}[htbp]
\center{
\includegraphics[scale=0.5]{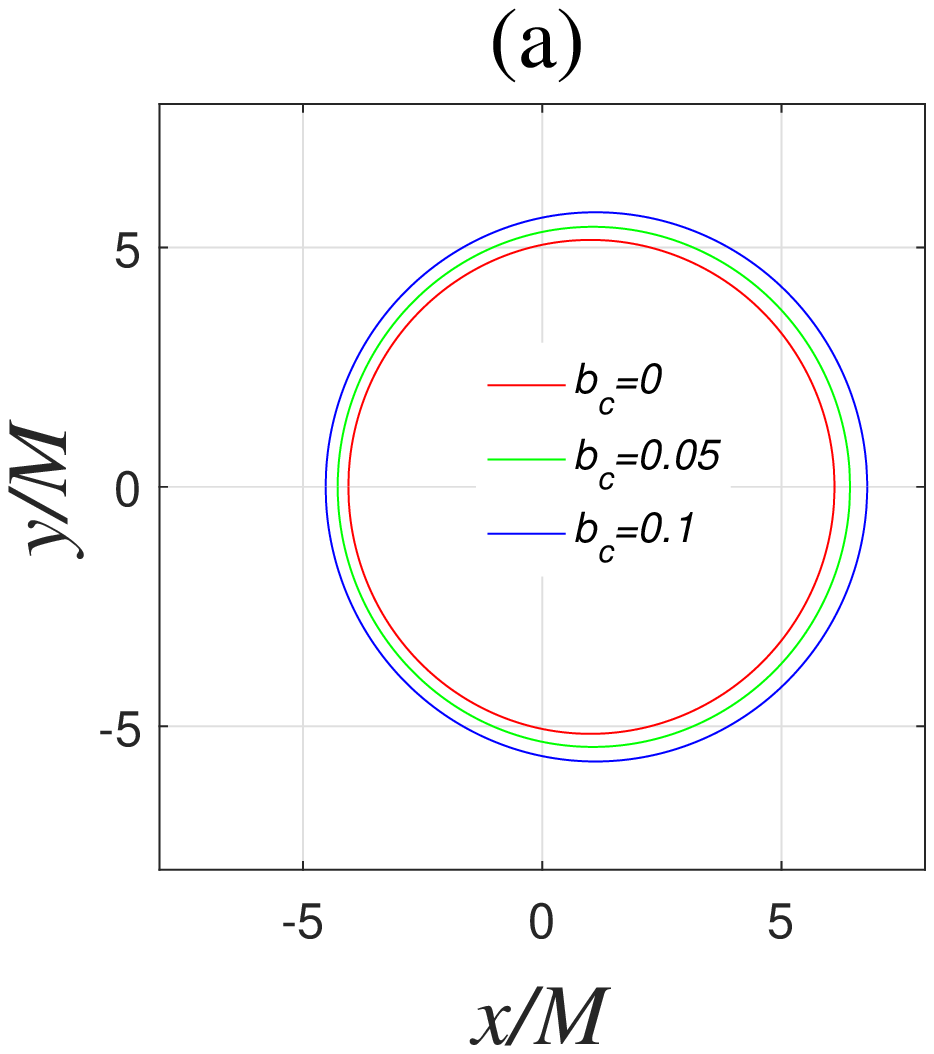}
\includegraphics[scale=0.5]{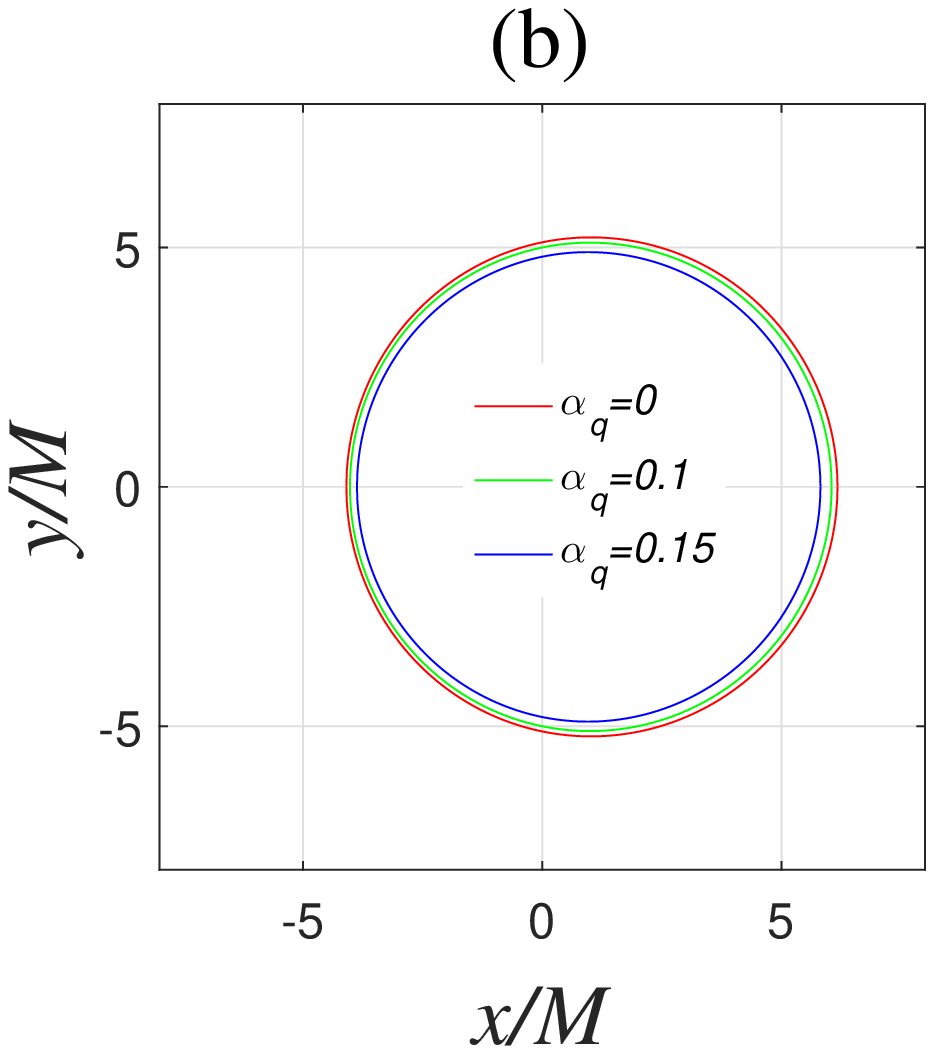}
\includegraphics[scale=0.5]{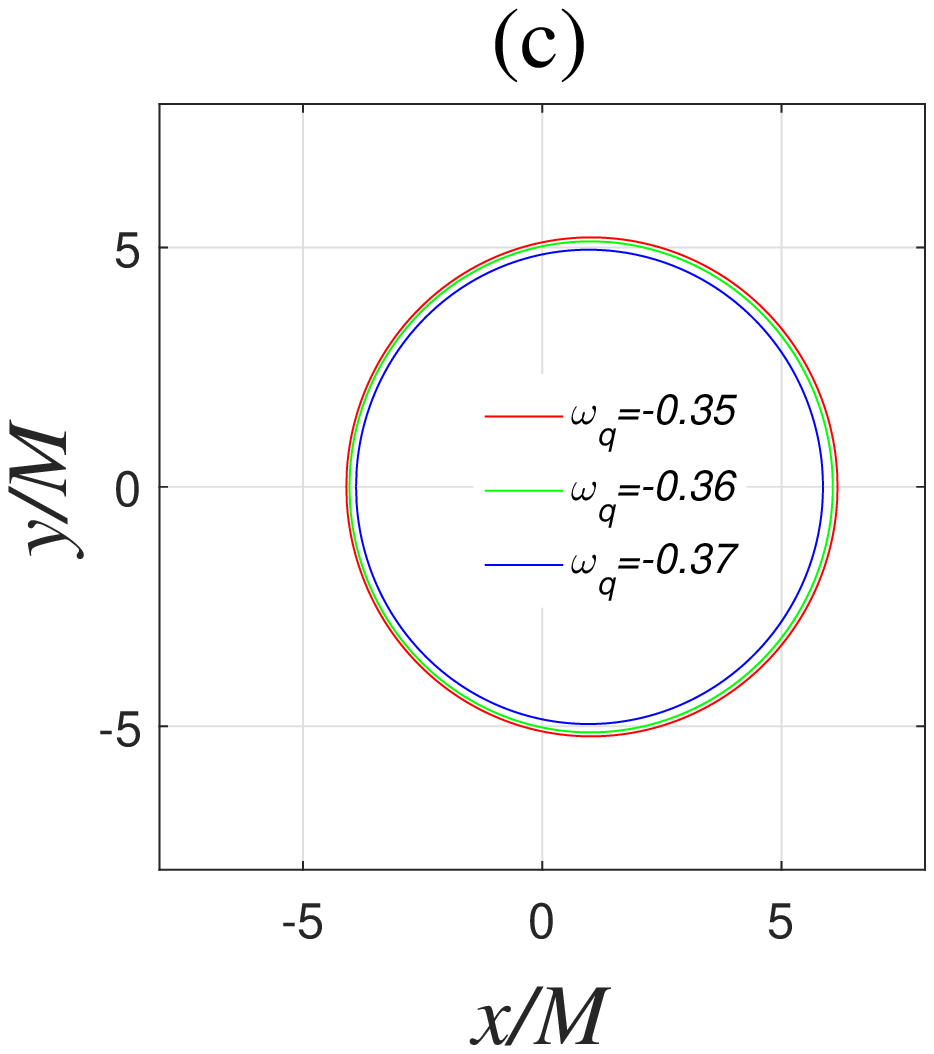}
\includegraphics[scale=0.5]{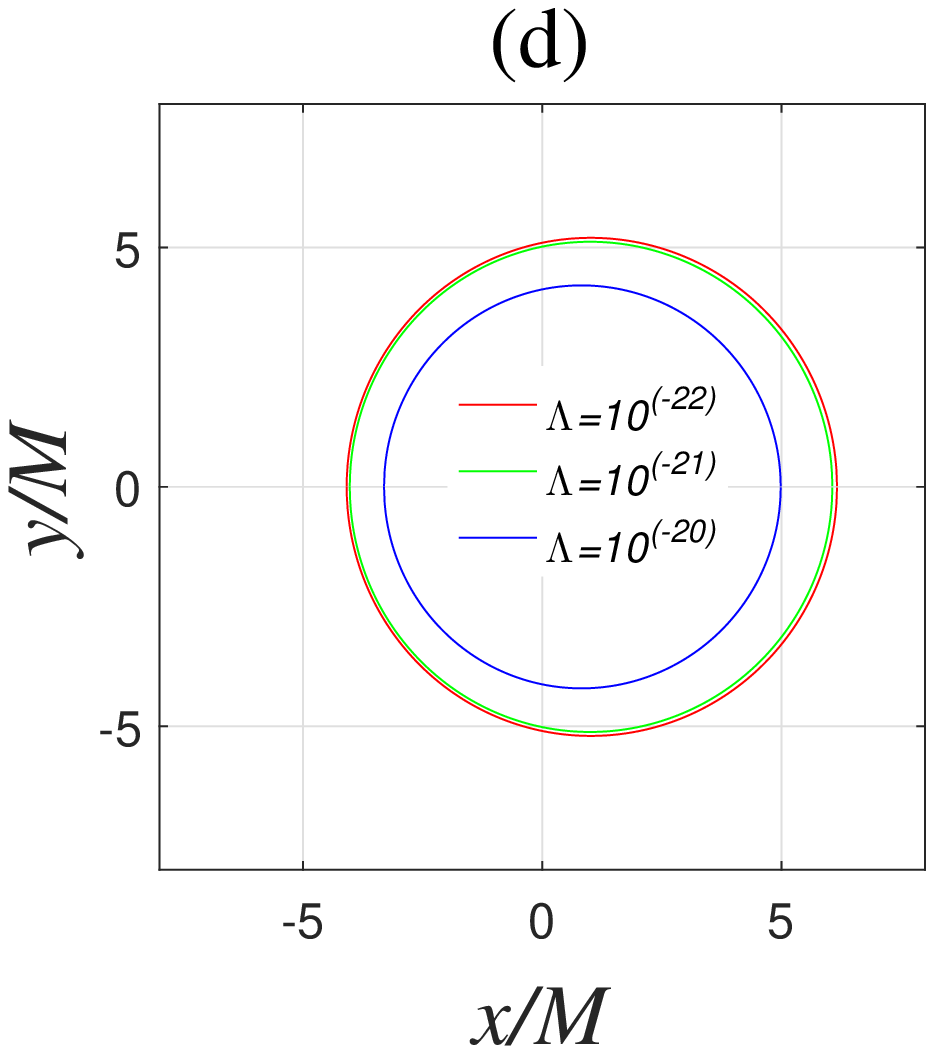}
\caption{Black hole shadows for different parameters. The parameters in these panels are $a=0.5$, $\theta_0=90^{\circ}$, and $Q=0.2$.
(a) $\alpha_q=0.01$, $\omega_q=-0.35$ and $\Lambda=1.02\times10^{-26}$.
(b) $b_{c}=0.01$, $\omega_q=-0.35$ and $\Lambda=1.02\times10^{-26}$.
(c) $b_{c}=0.01$, $\alpha_q=0.01$ and $\Lambda=1.02\times10^{-26}$.
(d) $b_{c}=0.01$, $\alpha_q=0.01$ and $\omega_q=-0.35$.}
\label{Fig6}}
\end{figure*}

\begin{figure*}[htbp]
\center{
\includegraphics[scale=0.5]{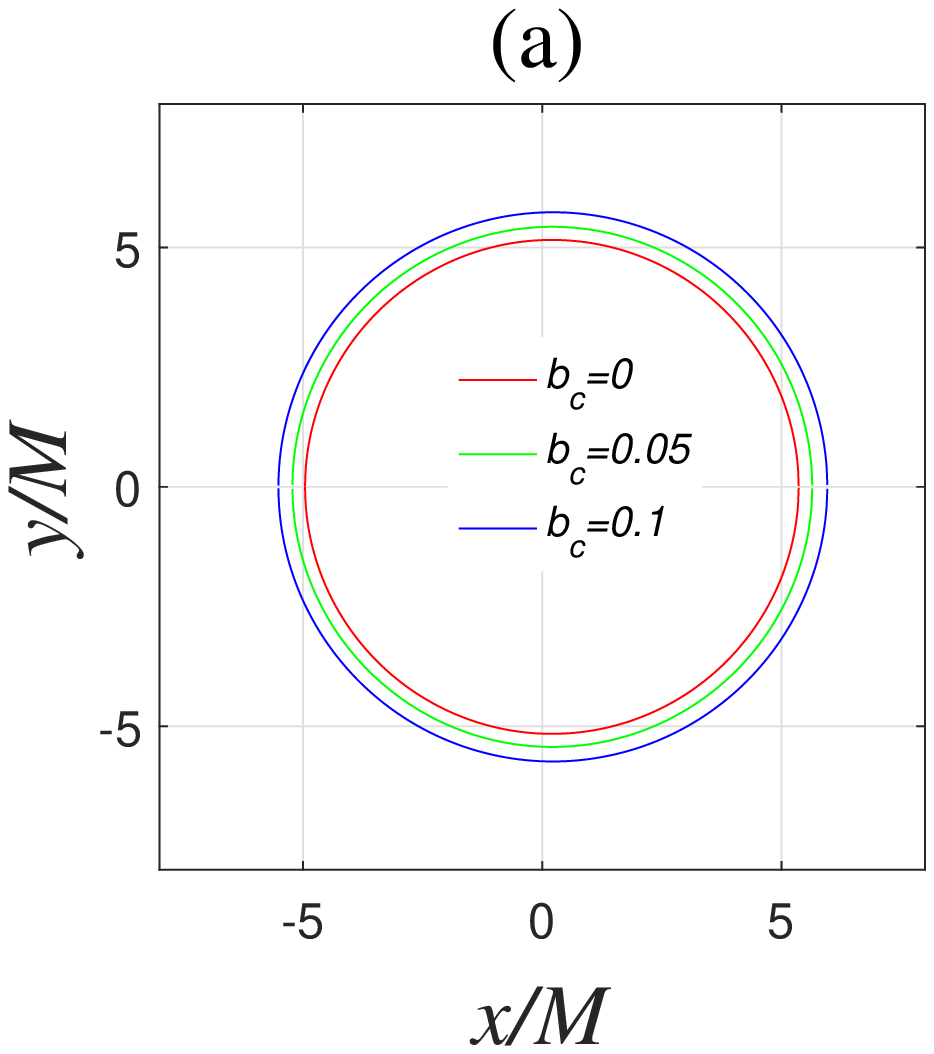}
\includegraphics[scale=0.5]{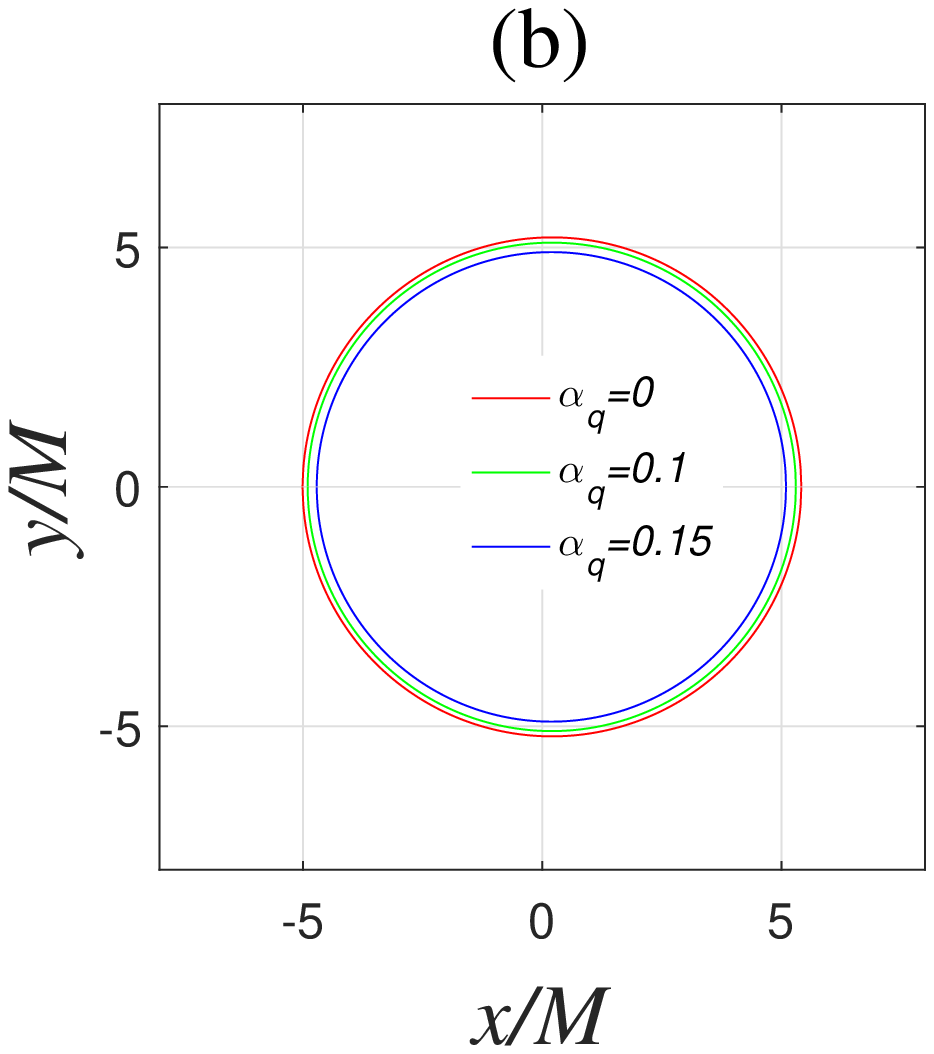}
\includegraphics[scale=0.5]{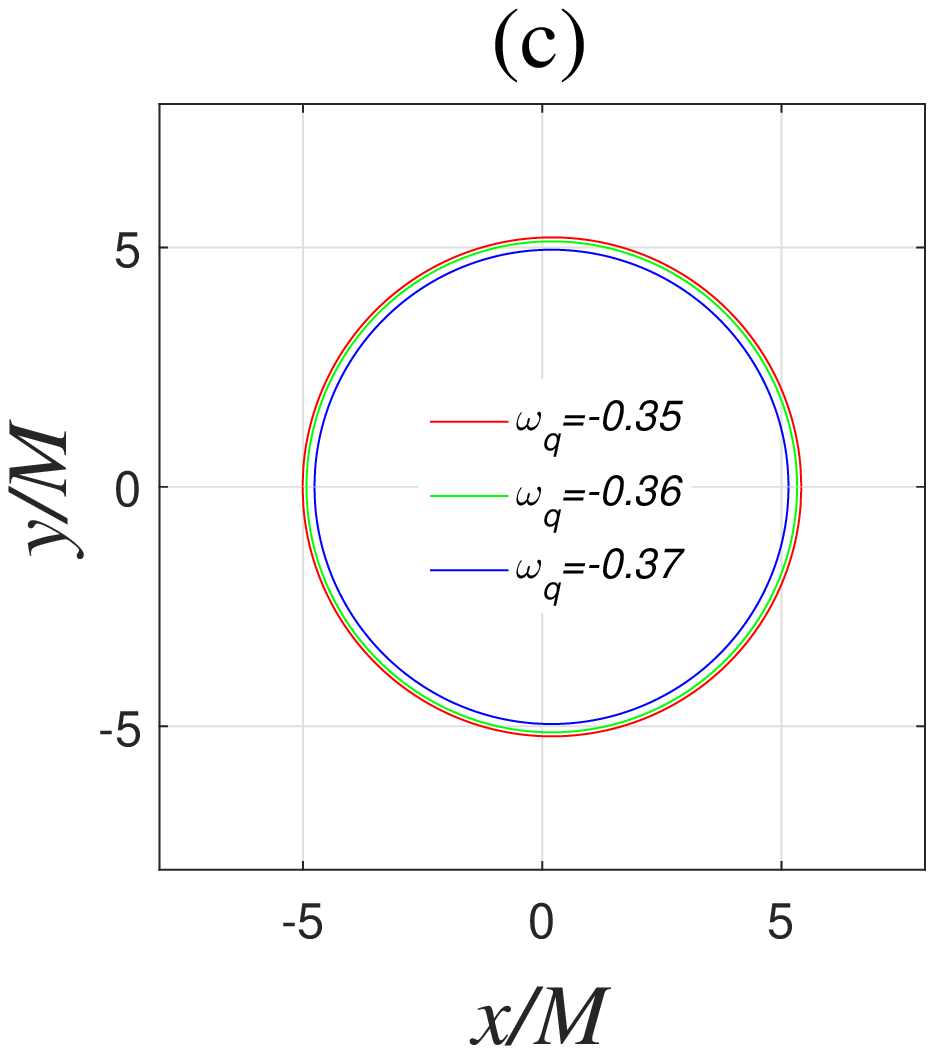}
\includegraphics[scale=0.5]{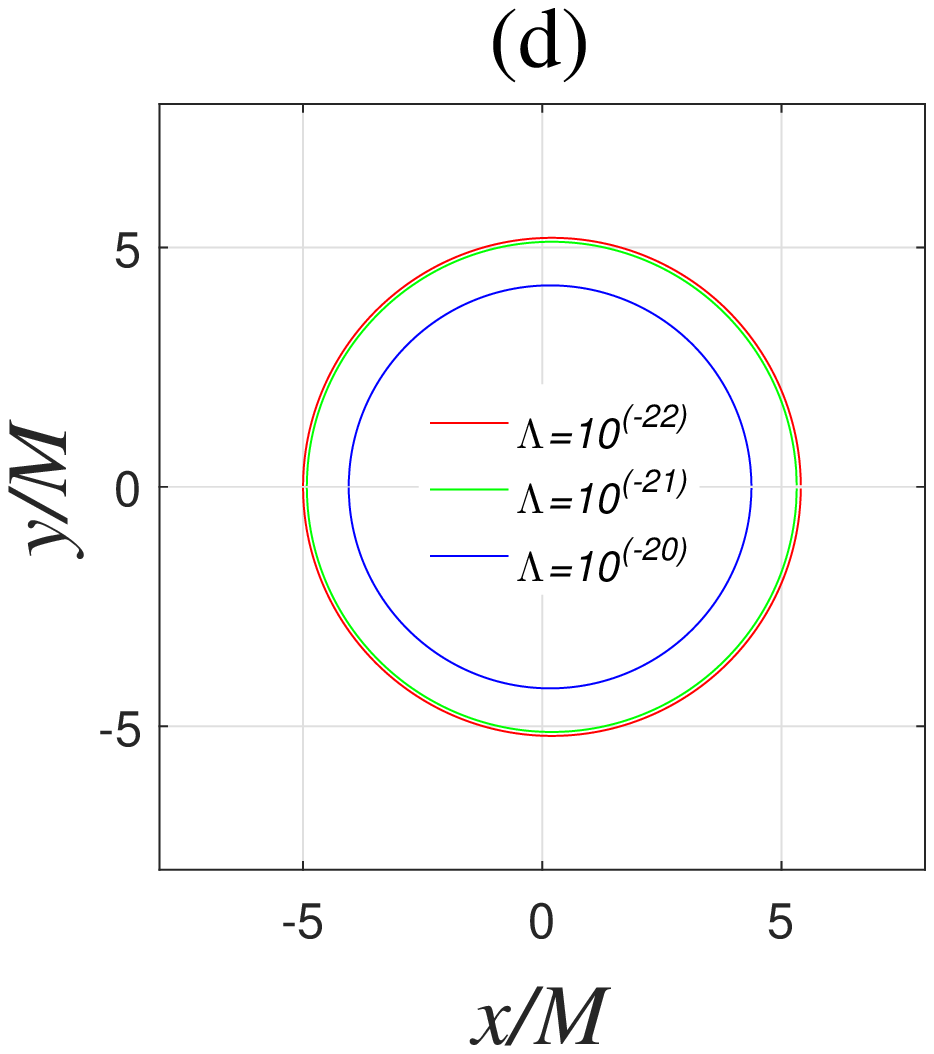}
\caption{Black hole shadows for several combinations of the parameters. The parameters in these panels are $a=0.1$, $\theta_0=90^{\circ}$, and $Q=0.2$.
(a) $\alpha_q=0.01$, $\omega_q=-0.35$ and $\Lambda=1.02\times10^{-26}$.
(b) $b_{c}=0.01$, $\omega_q=-0.35$ and $\Lambda=1.02\times10^{-26}$.
(c) $b_{c}=0.01$, $\alpha_q=0.01$ and $\Lambda=1.02\times10^{-26}$.
(d) $b_{c}=0.01$, $\alpha_q=0.01$ and $\omega_q=-0.35$.}\label{Fig6}}
\end{figure*}

\begin{figure*}[htbp]
\center{
\includegraphics[scale=0.5]{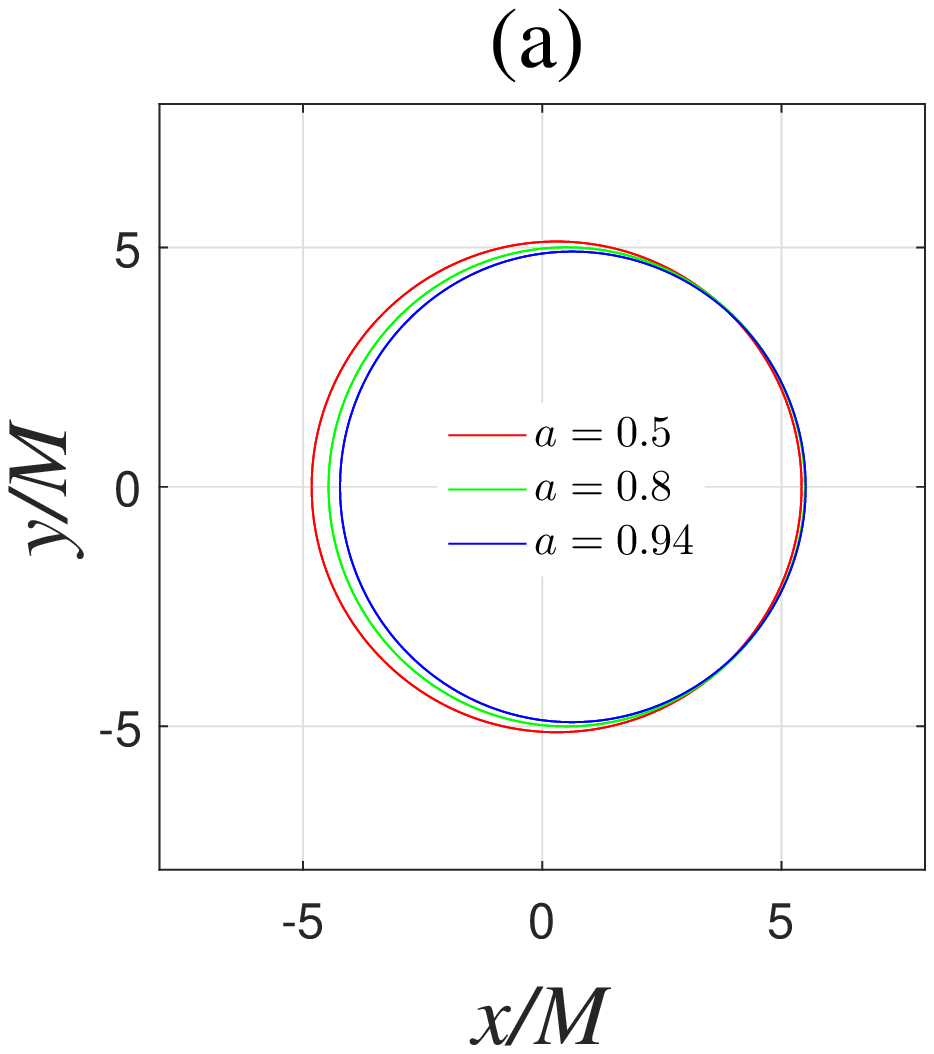}
\includegraphics[scale=0.5]{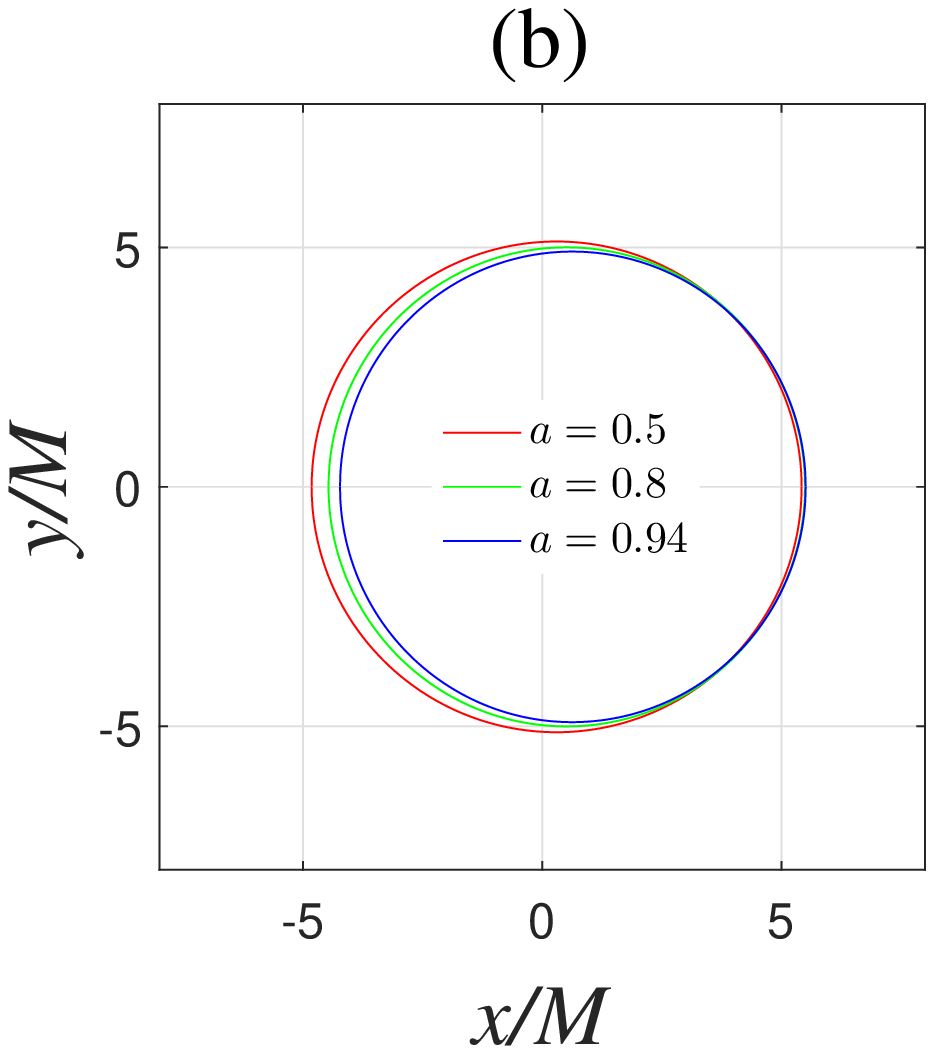}
\includegraphics[scale=0.5]{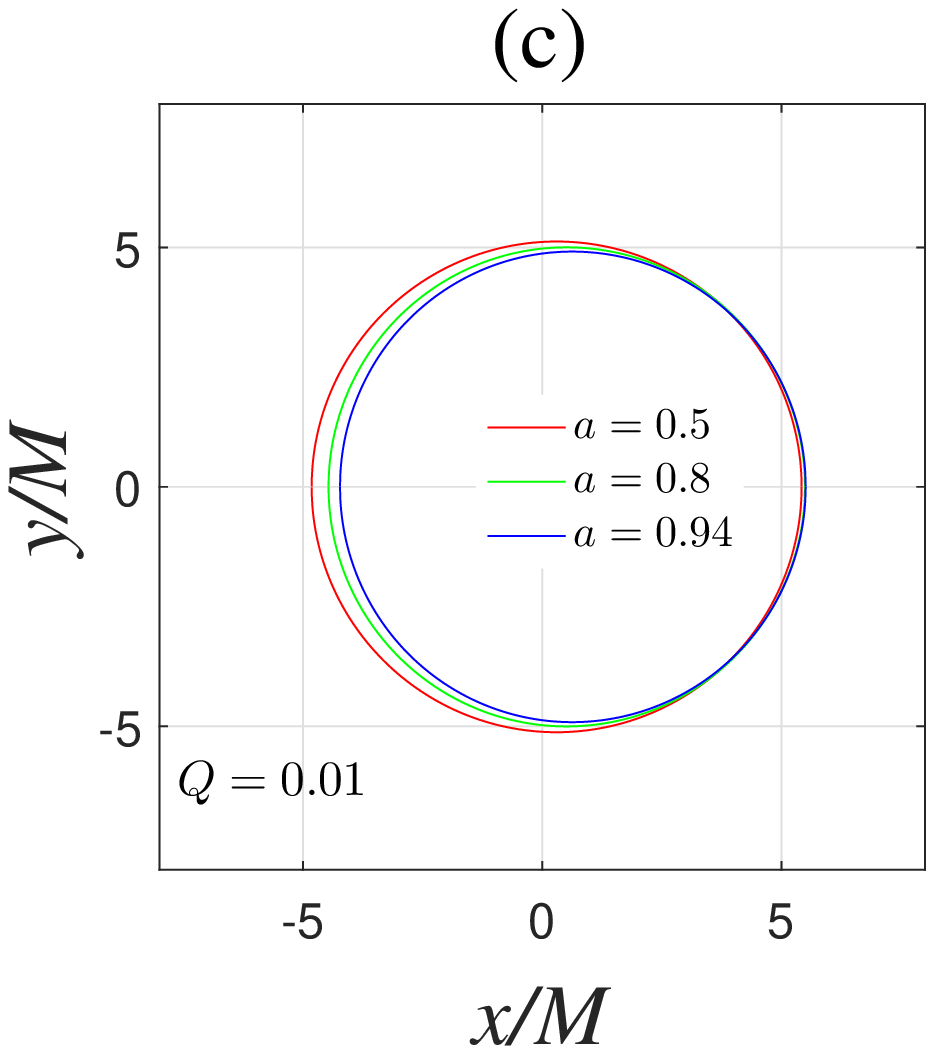}
\includegraphics[scale=0.5]{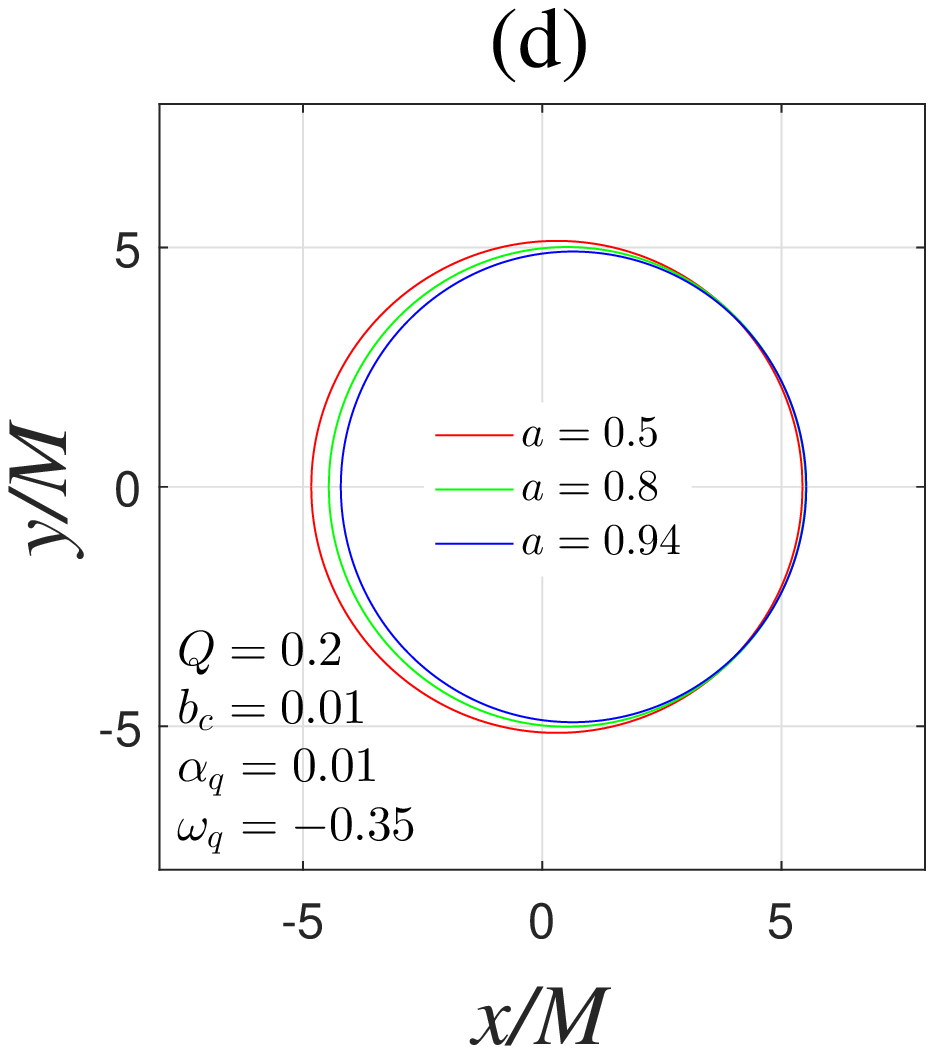}
\caption{Shadows of four black holes. The parameters in these panels are $\theta_0=17^{\circ}$ and $\Lambda=1.02\times10^{-26}$.
(a) Kerr black hole. (b) Kerr-de Sitter (KdS)  black hole. (c) KNdS black hole with $b_c=\alpha_q=0$. (d) KNdS black hole with $b_c\neq0$ and $\alpha_q\neq0$. }\label{Fig6}}
\end{figure*}

\begin{figure*}[htbp]
\center{
\includegraphics[scale=0.3]{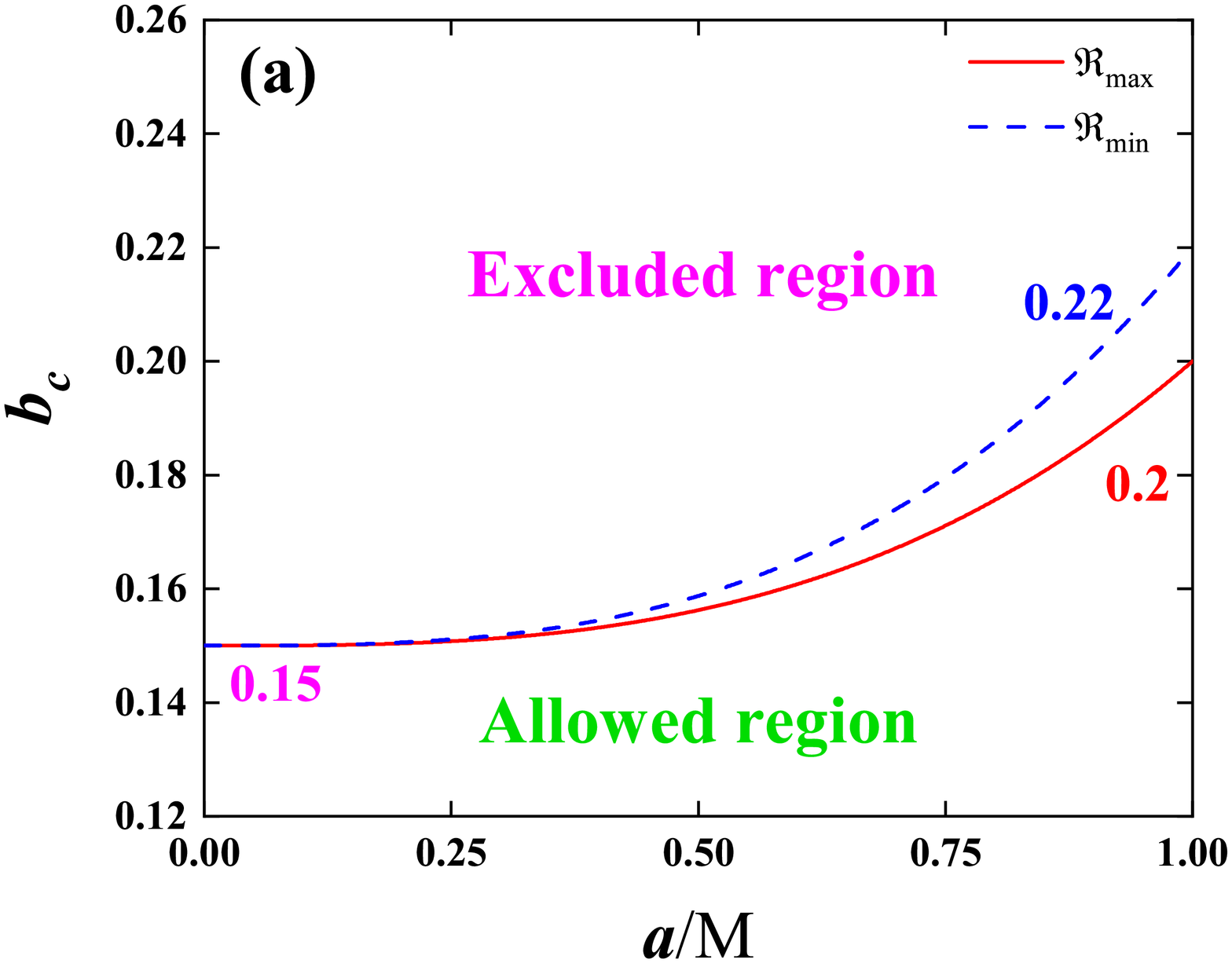}
\includegraphics[scale=0.3]{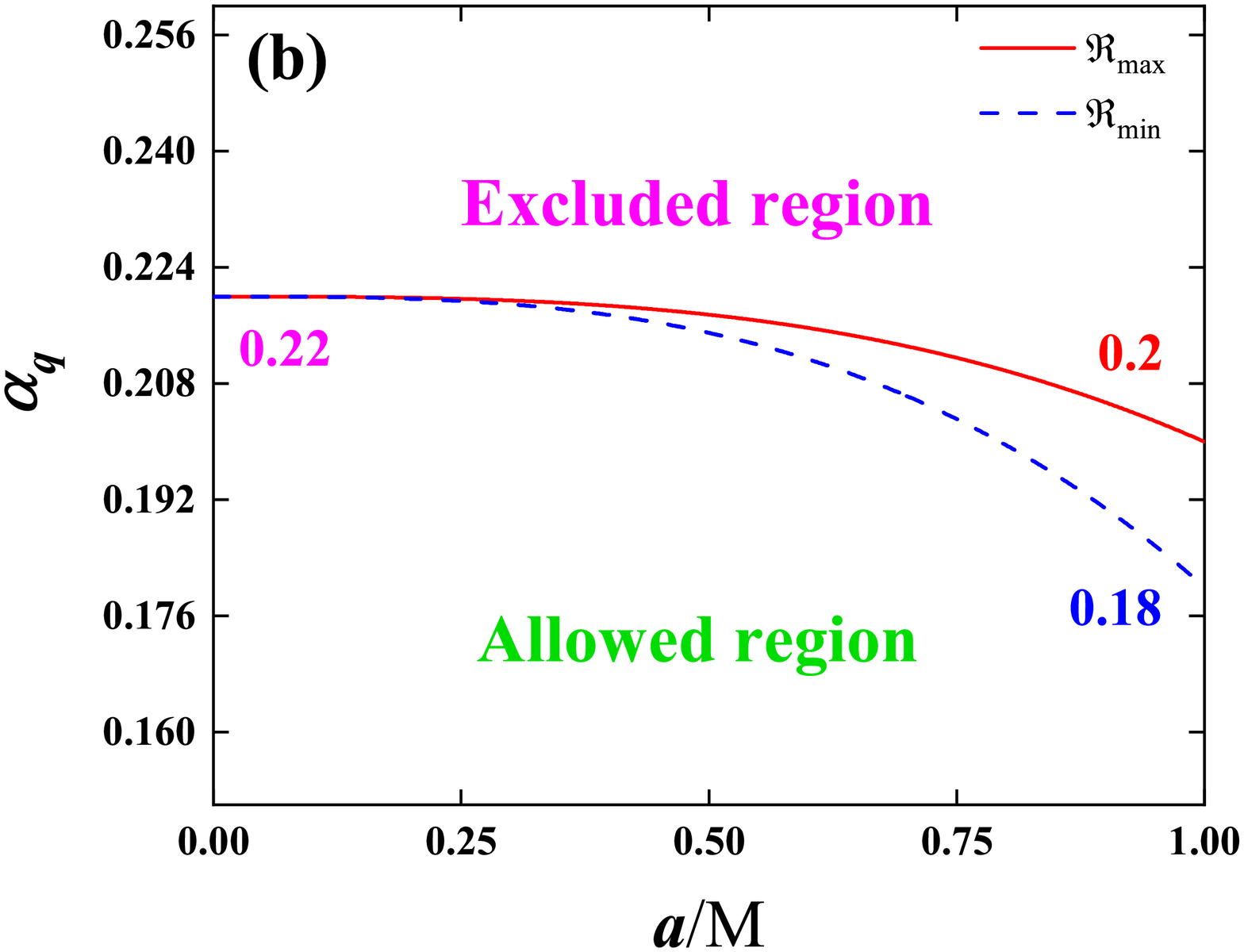}
\includegraphics[scale=0.3]{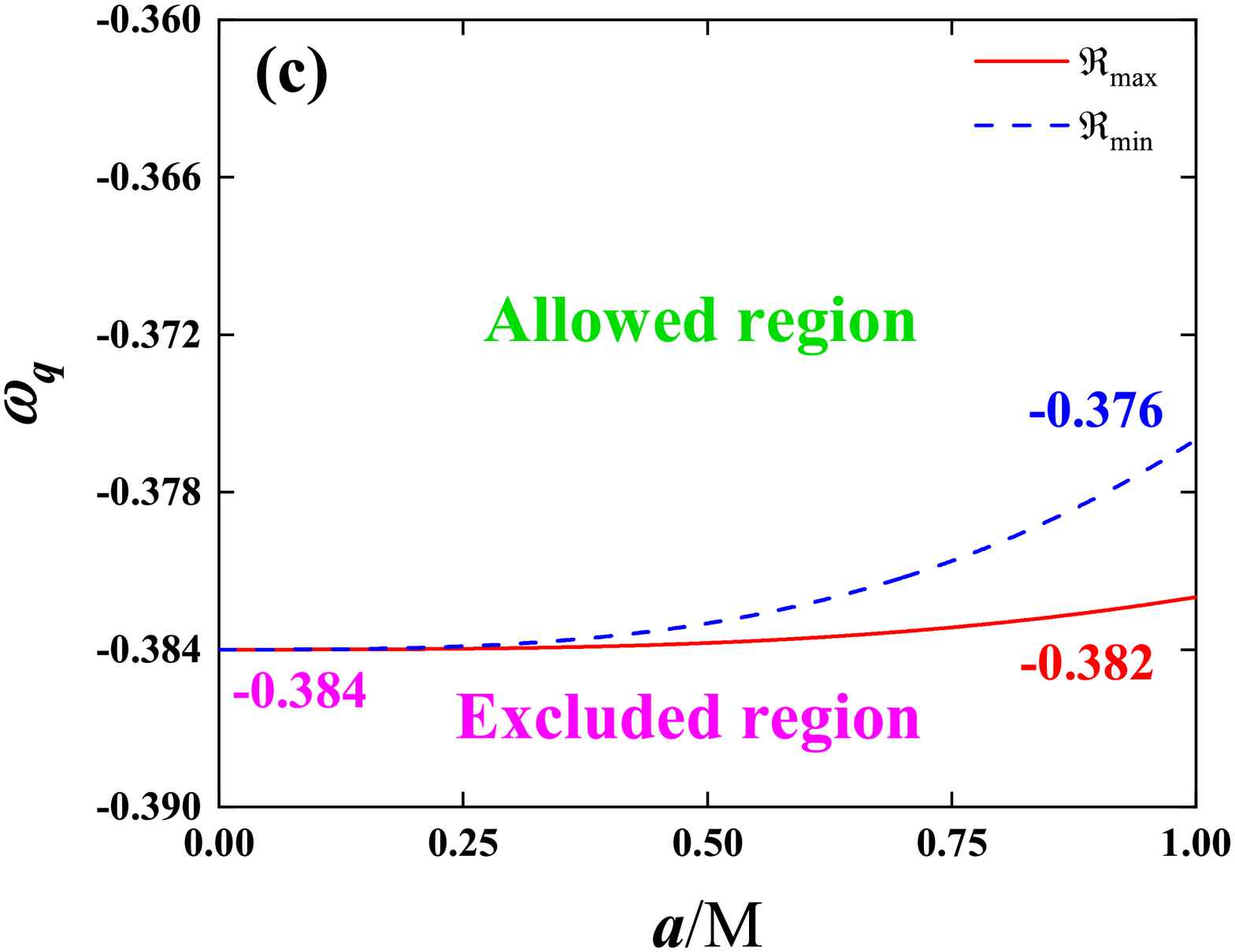}
\includegraphics[scale=0.3]{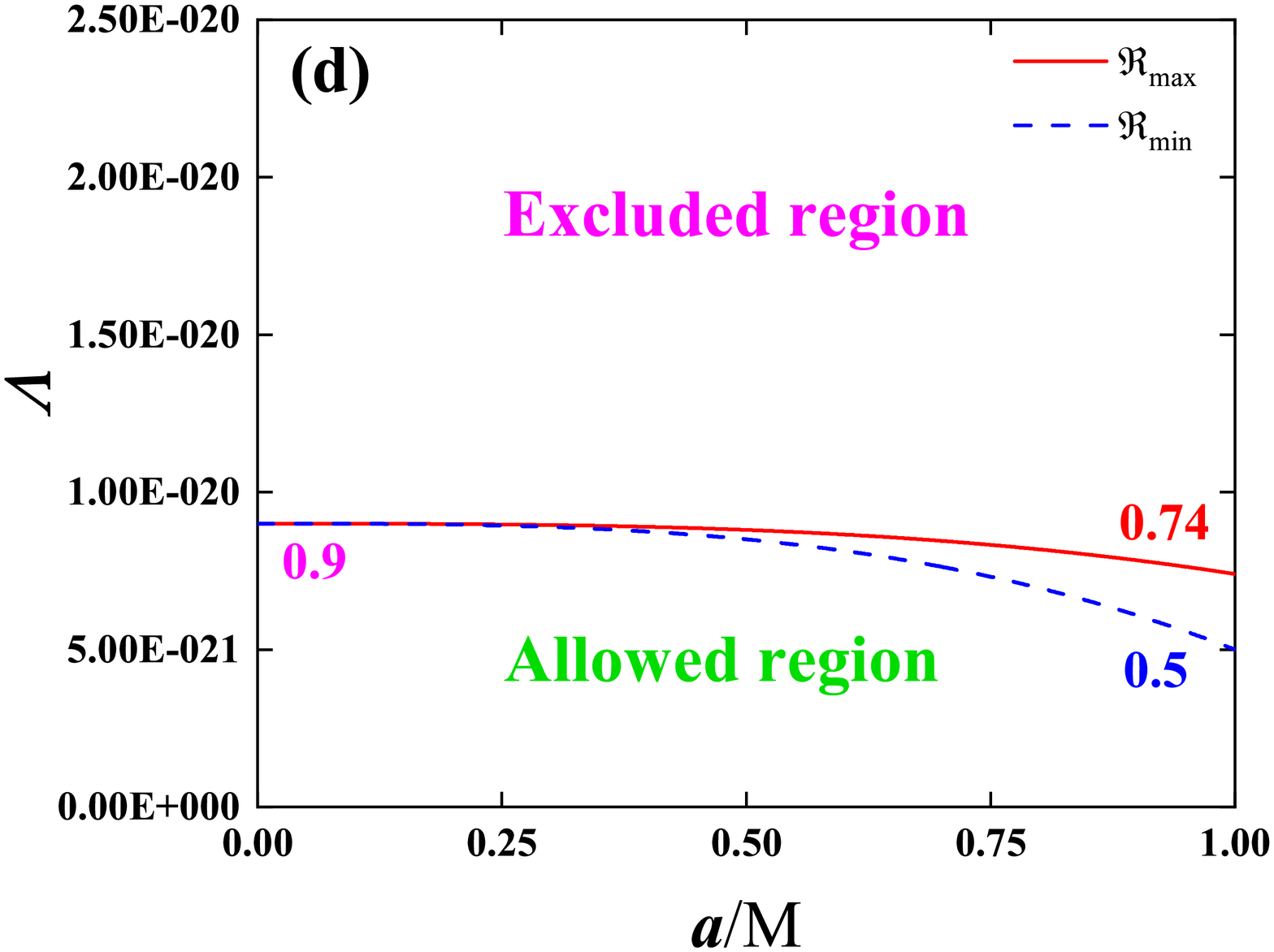}
\caption{Constraints of the parameters. The cloud strings and quintessence parameters in the ranges of  $0\leq b_c<0.15$, $0\leq\alpha_{q}<0.18$, $-0.376<\omega_{q}<-1/3$, and
$0\leq\Lambda<5\times10^{-21}$ are allowed. The cosmological constant is constrained in the range of $\Lambda<5\times10^{-47}m^{-2}$. The parameters in these panels are $\theta_0=17^{\circ}$ and $Q=0.2$.
(a) $\alpha_q=0.01$, $\omega_q=-0.35$ and $\Lambda=1.02\times10^{-26}$. The allowed region of $b_c$ is down the red solid curve.
(b) $b_{c}=0.01$, $\omega_q=-0.35$ and $\Lambda=1.02\times10^{-26}$. The allowed region of $\alpha_q$ is down the blue solid curve.
(c) $b_{c}=0.01$, $\alpha_q=0.01$ and $\Lambda=1.02\times10^{-26}$. The allowed region of $\omega_q$ is over the blue solid curve.
(d) $b_{c}=0.01$, $\alpha_q=0.01$ and $\omega_q=-0.35$. The allowed region of $\Lambda$ is down the blue solid curve.}
\label{Fig7}}
\end{figure*}

\end{document}